\documentstyle[graphicx,amsmath,amssymb,psfig,epsfig,multicols]{mn}  

\title{Measuring dark energy properties with 3D cosmic shear}
\author[A. F. Heavens, T. D. Kitching,
  A. N. Taylor]{A. F. Heavens,
  T. D. Kitching,
  A. N. Taylor\\ 
SUPA\thanks{The Scottish Universities Physics Alliance}, Institute for
Astronomy, University of Edinburgh, Royal Observatory, Blackford Hill,
Edinburgh, EH9 3HJ,UK\\
email: afh@roe.ac.uk, tdk@roe.ac.uk, ant@roe.ac.uk}

\newcommand{\be}{\begin{equation}}  \newcommand{\ee}{\end{equation}}
  \newcommand{\ba}{\begin{eqnarray}}
\newcommand{\ea}{\end{eqnarray}}

 \newcommand{\bg}{{\bf g}}  
  
  \newcommand{\br}{{\bf r}}

\newcommand{\hn}{{\hat{\bf n}}}

\newcommand{\bell}{{\mbox{\boldmath{$\ell$}}}}
\newcommand{\sbell}{{\mbox{\tiny\boldmath{$\ell$}}}}
\newcommand{\stp}{\sqrt{{2\over \pi}}}

\newcommand{\edth}{\,\eth\,}

\def\gs{\mathrel{\raise1.16pt\hbox{$>$}\kern-7.0pt %
\lower3.06pt\hbox{{$\scriptstyle \sim$}}}}         %
\def\ls{\mathrel{\raise1.16pt\hbox{$<$}\kern-7.0pt %
\lower3.06pt\hbox{{$\scriptstyle \sim$}}}}         %

\begin{document}

\maketitle

\begin{abstract}
We present parameter estimation forecasts for present and future 3D
cosmic shear surveys.  We demonstrate in particular that, in
conjunction with results from cosmic microwave background (CMB)
experiments, the properties of dark energy can be estimated with
very high precision with large-scale, fully 3D weak lensing surveys.
In particular, a 5-band, 10,000 square degree ground-based survey of
galaxies to a median redshift of $z_m=0.7$ could achieve $1$-$\sigma$
marginal statistical errors, in combination with the 
constraints expected from the CMB Planck Surveyor, of $\Delta
w_0=0.108$ and $\Delta w_a=0.099$. We 
parameterize the redshift evolution of $w$ by $w(a)=w_0+w_a(1-a)$
where $a$ is the scale factor. Such a survey is achievable with a
wide-field camera on a $4$ metre class telescope. The error on the
value of $w$ at an intermediate pivot redshift of $z=0.368$ is
constrained to $\Delta w(z=0.368)=0.0175$. We compare and combine
the 3D weak lensing constraints with the cosmological and dark
energy parameters measured from planned Baryon Acoustic Oscillation
(BAO) and supernova Type Ia experiments, and find that 3D weak
lensing significantly improves the marginalized errors on $w_0$ and
$w_a$ in combination, and provides constraints on $w(z)$ at a unique
redshift through the lensing effect.  A combination of 3D weak
lensing, CMB and BAO experiments could achieve $\Delta w_0=0.037$
and $\Delta w_a=0.099$. We also show how our results can be scaled
to other telescopes and survey designs. Fully 3D weak shear analysis
avoids the loss of information inherent in tomographic binning, and
we also show that the sensitivity to systematic errors in
photometric redshift is much less. In conjunction with the fact that
the physics of lensing is very soundly based, the analysis here
demonstrates that deep, wide-angle 3D weak lensing surveys are
extremely promising for measuring dark energy properties.
\end{abstract}

\begin{keywords}
cosmology: observations - gravitational lensing - large scale
structure, galaxies: formation
\end{keywords}

\section{Introduction}
Our knowledge of cosmology has advanced considerably in recent
years. Led by detailed measurements of the microwave background
radiation and large-scale structure, many of the most important
cosmological parameters are now known with good accuracy. This
advance has come about principally through the all-sky maps of the
microwave sky taken with the Wilkinson Microwave Anisotropy Probe
(WMAP) (Bennett et al., 2003), supplemented by higher-resolution
observations of the Arcminute Cosmology Bolometer Array Receiver
(ACBAR) and Cosmic Background Imager (CBI) (Kuo et al., 2004; Pearson
et al., 2003). When
combined with large-scale structure information from the
Anglo-Australian 2 degree field galaxy redshift survey (2dFGRS)
(Colless et al., 2001; Percival et al., 2001), the Lyman-$\alpha$ forest
(Croft et al., 2002; Gnedin and Hamilton, 2002), and measurements of galaxy bias
(Verde et al., 2002), the data establish the concordance model of an
accelerating Universe dominated by dark energy and dark matter
(Spergel et al., 2003; Spergel et al., 2006).  The acceleration of the Universe is
also apparent in observations of distant supernovae (e.g.
Riess et al., 2000).  The determination of the density, baryon content
and expansion rate of the Universe shifts the major unanswered
questions in cosmology to the nature of the dark matter and dark
energy. Dark energy in particular can be probed through its
cosmological effects on the distance-redshift relation and the
growth rate of structure. 

The question of the precise nature of the dark energy is a
far-reaching one. We use the simplest phenomenological model
of dark energy by 
parameterizing the equation of state of the vacuum,
\begin{equation}
w\equiv p/(\rho c^2),
\end{equation}
where $\rho$ is its energy-density and $p$ is the
dark-energy/vacuum-pressure. $w=w(a)$ may vary with scale factor. 
If it is found with high precision that $w=-1$,
then the dark energy cannot be distinguished with large-scale
measurements from a modification to the gravity law along the lines
suggested by Einstein with the cosmological constant.  If, however,
it can be established with a degree of certainty that $w$ differs
from $-1$ at any redshift, then it cannot be associated with such a
change to the 
gravity law, and is most naturally accounted for by a new field.
This would be an extremely important discovery, and the
time-evolution of the field would be a useful constraint on models.
Some possibilities exist in the literature, such as those proposed
by Ratra and Peebles (1988), but none is a clear favourite candidate.

Weak lensing is a very attractive proposition for studying dark
energy, as it is sensitive to both of these effects, and, equally
importantly, the physics of weak lensing is well understood.  A key
part of this is that it is sensitive to the distribution of matter
in the Universe, regardless of its form.  Furthermore, since weak
lensing analysis can be done in a way which is either dependent on
the distance-redshift relation alone (see e.g. Taylor et al., 2006;
Jain \& Taylor, 2003) or on both the distance-redshift relation 
and the growth factor (this paper), it can in principle distinguish
between modified gravity models and dark energy models.

The main lesson of the field of microwave background astronomy is
that with well-understood physics, robust results can be obtained
with high precision.  Weak lensing observations are, however, a
technical challenge, as the imaging requirements are severe.  Thus,
it has only been in the last five years or so that the first
measurements of cosmic shear have appeared
(Bacon, Refregier and Ellis, 2000; Kaiser, Wilson and Luppino, 2000;
van Waerbeke et al., 2000; Wittman et al., 2000). Weak lensing measurements
to date have concentrated on obtaining the matter density parameter
$\Omega_m$ and the amplitude of mass density fluctuations
(Hoekstra, Yee and Gladders, 2002; Jarvis et al., 2003; Rhodes et al.,
2004; Heymans et al., 2004; Hoekstra et al., 2006; Semboloni et al., 2006).
More ambitiously, weak lensing observations have started to put
constraints on the equation of state of dark energy
(Jarvis et al., 2005; Semboloni at al., 2006).  Theoretically, the prospects for
determining dark energy properties (specifically its equation of
state $w$) using weak lensing have been explored in a number of
papers (e.g. Taylor et al., 2006; Hu and Tegmark, 1999; Huterer, 2002;
Heavens, 2003; Refregier, 2003; Simon, King and Schneider, 2004;
Takada and Jain, 2004; Song and Knox, 2004; Ishak et al., 2004; Ishak, 2005). 
The prospects for determining $w$ as a function of redshift $z$ are
markedly improved when 3D information on the individual lensed
sources is available.  Source distances could come from
spectroscopic redshifts, but given the depth and the sky area
required, they are more likely to be estimated from photometric
redshifts.  With 3D information, the lensing pattern can be analyzed
in shells at different distances (e.g. Hu, 1999; Hu and Jain, 2004;
Ishak, 2005), or by analyzing the shear pattern as
a fully three-dimensional field (Heavens, 2003).  It is the latter
possibility which we investigate in this paper.  The statistical
properties of the shear pattern are influenced by many cosmological
parameters, including $w(z)$.  In this paper we extend the analysis of
Heavens (2003) to small-angle surveys as well as computing the expected
marginal errors on $w$ (and its evolution), using a Fisher matrix
approach.  We investigate issues of depth vs area, and the number of
photometric bands which should be used, to determine the dark energy
properties as accurately as possible.  The main focus of the paper
is in computing the expected statistical errors, but we do consider
the impact of some systematics (Ishak et al., 2004; Bernstein, 2005;
Huterer et al., 2005).

The layout of the paper is as follows:  in Section \ref{s:method} we
detail the transform method used and compute the covariance matrix
of the transform coefficients; in Section \ref{s:params} we outline
how the expected statistical errors on parameters are calculated; in
Section \ref{s:Errors} we present the survey design and how we can scale to
other surveys and in Section \ref{Optimisation For a Wide-Field
  Lensing Survey} we  present an
optimization of survey design and the parameter errors; in Section
\ref{Parameter Forecasts} we consider the synergy of 3D weak lensing
with other dark energy probes and discuss future surveys and finally
we give our conclusions in Section \ref{Conclusions}. 

\section{Method}\label{s:method}
\subsection{Transformation of scalar and shear
fields}\label{ss:transforms}
The observable quantities we use are the estimates of the shear
field at locations in three dimensions.  The estimates of the
complex shear come from the shape and orientation of galaxies, where
the radial distance is obtained approximately by using photometric
redshift estimates obtained from observations through several or
many filters.

In a previous paper (Heavens, 2003) we introduced the idea of 3D
weak lensing analysis in harmonic space as a statistical tool. In
Castro, Heavens and Kitching (2005), we developed the subject formally and found the
power spectrum of 3D weak lensing shear. In this paper we consider the
flat-sky limit including the non-linear evolution of the power
spectrum. We consider a transform of
the 3D shear field in spin-weight spherical harmonics and spherical
Bessel functions. This is a very natural expansion for the shear
field, as the complex shear $\gamma$ is a spin-weight 2 object, as
are the spin-weight 2 spherical harmonics: under a local rotation of
the coordinate system by angle $\psi$, $\gamma$ changes to $\gamma
e^{2i\psi}$. The spherical harmonic transform of a spin-weight $s$
field $_sf(\br)$ is defined here by
\begin{equation}
_sf_{\ell m}(k) \equiv \sqrt{2\over \pi} \, \int\,d^3\br _sf(\br)\,
k j_\ell(kr)\,_sY_\ell^{m*}(\hn)
\end{equation}
where $j_\ell(z)$ is a spherical Bessel function, $_sY_\ell^m$ a
spin-weight $s$ spherical harmonic, $k$ is a radial wavenumber,
$\ell$ is a positive integer, $m=-\ell, \ldots \ell$ and $\hn$
represents the direction $\theta$, $\varphi$.  For $s=0$ the
spin-weight spherical harmonics are the usual spherical harmonics
$Y_\ell^m$, and this is the appropriate spherical expansion of a
scalar field.  Note the presence here of a benign factor of $k$, to
agree with the notation of Castro, Heavens and Kitching (2005).  The motivation for
using spherical coordinates is manyfold: firstly the selection
function for a survey can often be separated into an angular (sky
coverage) part and a radial component; secondly the errors in
photometric redshifts introduce purely radial errors in the
positions of the source galaxies; thirdly, in the Born
approximation, the lensing effect is an integral effect along the
(radial) line of sight. The motivation in flat space for using
products of spherical Bessel functions and spherical harmonics is
that, as eigenfunctions of the Laplacian operator, it is easy to
relate the expansion coefficients of the gravitational potential to
those of the density field.  Similar considerations led Heavens and
Taylor (1995)(see also Fisher et al., 1994; Tadros et al., 1999;
Percival et al., 2004) to expand the 
large-scale structure of galaxies in spherical Bessel functions and
spherical harmonics.  Since cosmic shear depends on the
gravitational potential, the use of this basis allows us to relate
the expansion of the shear field to the expansion of the mass
density field. The properties of the latter depend in a calculable
way on cosmological parameters, so this opens up the possibility of
using 3D weak shear to estimate these quantities.

For surveys with large opening angles on the sky, a full expansion
in spherical Bessel functions and spherical harmonics is the natural
choice. Such an expansion is generally applicable, but for
small-angle surveys whose signal is dominated by high-$\ell$ modes,
the spherical harmonics are cumbersome and their accurate
computation can present problems. For such surveys, we can
approximate the spherical harmonics as sums of exponentials, as
detailed in Appendix A of Santos et al. (2003).

For this paper we use the flat-sky expansion, which for a scalar
($s=0$) field reads
\begin{equation}
f(k,{\bell}) \equiv \sqrt{2\over\pi} \, \int\,d^3\br f(\br) k
j_\ell(kr) \exp(-i\bell\cdot\btheta),
\label{flatsky}
\end{equation}
where $\bell$ is a 2D angular wavenumber and $k$ a radial
wavenumber. In the spherical Bessel function, $\ell = |\bell|$;
$\ell$ is necessarily an integer, but we assume that $\ell \gg 1$ so
that enforcing integer $\ell$ is a minor approximation. Note that we
are performing a full 3D expansion of the shear field and assume a
flat Universe except where indicated. An alternative approach to include at
least some 3D information is what is referred to as tomography,
where the shear pattern of galaxies is analyzed in shells, based on
their photometric redshifts (Hu, 1999; Hu, 2002; Jain and Taylor,
2003; Takada and White, 2004). It is
however evident that the binning process loses at least some
information, and it is not necessary.

The inverse transform in the flat-sky approximation is
\begin{equation}
f(\br) = \stp \int {d^2\bell\over (2\pi)^2} dk\, k  j_\ell(kr)
\exp(i\bell.\btheta)f(k,\bell).
\end{equation}
The coefficients of the expansion in the two systems are related by
generalization of equation (A13) in Santos et al. (2003):
\begin{equation}
f(k,\bell) = \sqrt{{2\pi\over \ell}}\sum_m i^{-m}f_{\ell m}(k)
\exp(im\phi_{\sbell})
\end{equation}
where the small survey is centred at the pole of the coordinate
system, and the 2D transverse wavevector is $\bell =
(\ell\cos\phi_{\sbell},\ell\sin\phi_{\sbell})$.  The covariances of
the flat-sky coefficients are related to the power spectrum of $f$
by
\begin{equation}
\langle f(k,\bell)f^*(k',\bell')\rangle  = (2\pi)^2\,P_f(k)
\,\delta^D(k-k')\delta^D(\bell-\bell')
\end{equation}
where $\delta^D$ is the Dirac delta function.

Our plan is essentially to transform the components of the 3D shear
field to produce a set of transform coefficients as a function of
$(k,{\bf \bell}$).  These data will depend on cosmological
parameters, and can be used in a likelihood analysis to constrain
those parameters.

\subsubsection{Transformation of shear fields}
The weak lensing shear components we transform are $\gamma_1({\bf
r})$ and $\gamma_2({\bf r})$, which are related to the lensing
potential $\phi({\bf r})$ through (e.g. Bartelmann and Schneider, 2001) 
\begin{equation}
\gamma_1 = {1\over 2}\left(\phi_{11}-\phi_{22}\right);\qquad
\gamma_2 = \phi_{12}
\end{equation}
where $\phi_{ij}\equiv \partial^2 \phi/\partial \theta_i \partial
\theta_j$.  $\phi$ itself is dependent on cosmological parameters
through its relation to the mass density field (see section
\ref{ss:Gsec}). We will return to this dependence later.  For a
large-area survey, it is a measure of the shears with respect to axes
based on the spherical coordinate system, in which case the complex
shear $\gamma\equiv \gamma_1 + i\gamma_2$ is the second edth
derivative of $\phi$:
\begin{equation}
\gamma({\bf r}) = {1\over 2}\edth\edth\phi({\bf r}).
\end{equation}
(Castro, Heavens and Kitching, 2005). In the flat-sky limit, $\edth\rightarrow
-(\partial_x+i\partial_y)$, where the $\partial_{x,y}\equiv
\partial/\partial\theta_{x,y}$.  Expanding the lensing potential in
terms of spherical Bessel functions and exponential functions, as in
equation (\ref{flatsky}), we see that it is natural to expand the complex
shear field in terms of $\edth\edth\exp(-i\bell.\btheta) = \ell^2
X_\bell\exp(-i\bell.\btheta)$, where
\begin{equation}
X_\bell \equiv{(\bell_y^2-\bell_x^2)+2i\bell_x\bell_y\over \bell^2}.
\end{equation}
The $\bell^2$ in the denominator is included for convenience, so the
inverse transform kernel is just $\sqrt{2/\pi} k\,j_\ell(kr)
X_\bell^* \exp(i\bell.\btheta_g)$.

\subsubsection{Fiducial Cosmology}
An immediate issue to address is which radial coordinate to use in
the spherical Bessel function.  The observed quantities are the
estimated redshifts of the sources, and we need to do two things:
one is to translate these into radial distances; the second is to
account for the error in the estimation of the redshifts.  For the
former, we choose a fiducial set of cosmological parameters, to
define a transformation $r^0(z_p)$ from the photometric redshift
estimate $z_p$ to a radial coordinate $r^0$.  For this paper, we
choose as the fiducial model the concordance model (Spergel et al.,
2006) with $\Omega_m=0.27$,  $\Omega_b=0.04$,
$\Omega_v=0.73$, $\sigma_8=0.8$, $h=0.71$, $w_0=-1$ and $w_a=0$
where the variables are the matter, baryon, vacuum density
parameters, Hubble constant in units of $100\,$
km$\,$s$^{-1}\,$Mpc$^{-1}$ and the dark energy equation of state
parameters respectively. The equation of state of dark energy is
modelled in terms of scale factor $a$ by 
\begin{equation}
\label{DEeqstate}
w(a) = w + w_a(1-a)
\end{equation}
(Chevallier and Polarski, 2001; Linder, 2003) where $a(z)=(1+z)^{-1}$ is the cosmic
scale factor normalized to unity at the present epoch.

We also include the scalar spectral index
$n_s=1$ and its running $\alpha_n=0$. For the CMB Fisher
calculations, see Section \ref{Combining with other dark energy
experiments},
 we also include the tensor to
scalar ratio $r=0.01$ and the optical depth to the surface of last scattering $\tau=0.09$.

\subsection{Transformation}
The lensing potential is defined everywhere, but we sample it only
at the locations of galaxies, so it is natural to make a
transformation of this point process, summing over galaxies rather
than integrating over space.  Our estimate of the transform is thus
defined as
\begin{equation}
\hat\gamma(k,\bell) = \stp \sum_g \,\gamma(\br)k j_\ell(k
r_g^0)\exp(-i\bell.\btheta_g) W(r_g^0)
\end{equation}
where $W(r)$ is an arbitrary weight function, and $(r_g^0,\btheta_g)$
are the coordinates of galaxy $g$.

Note the appearance of two distances in the transform, $r$ and $r^0$
(at each galaxy $g$): the main application of this study is to
determine cosmological parameters, which affects the $r(z)$
relation. The shear field is the shear field at the {\em actual}
coordinate ${\bf r}_g$ of the galaxy, and this depends on the true
cosmological parameters, whereas the expansion (and weighting) is
done with the fiducial model parameters. This distinction was
neglected in Heavens (2003) and leads to an underestimate of the
errors on the dark energy equation of state in that paper; the error
estimates for the power spectrum in that paper are unaffected by
this error.

Writing the number density of source galaxies $n(\br)$ as the sum
of a set of delta functions, we see that
\begin{equation}
\hat\gamma(k,\bell) =\stp \int d^3\br\, n(\br)
\gamma(\br)\,k\,j_\ell(kr^0)\exp(-i\bell\cdot \btheta) W^0,
\label{Gammahatcont}
\end{equation}
where $W^0=W(r^0)$. Note that in the high-$\ell$ limit these are also
the (minus) coefficients of the expansion of the convergence field
$\kappa$ (Castro, Heavens and Kitching, 2005). This has an expectation
value which is obtained by replacing $n(\br)$ by the mean density of
the source galaxies, 
$\bar n(r)$. Here we assume that selection effects are uniform
across the survey so there is no angular dependence.  Thus the
$\hat\gamma$ are estimators of
\begin{equation}
\gamma(k,\bell) \equiv \stp \int d^3\br\, \bar n(r) \gamma(\br)\,k
j_\ell(kr^0)\exp(-i\bell.\btheta) W^0. \label{Gammar}
\end{equation}
The estimates will differ because of the discrete nature of the
galaxies, which leads to shot noise, the photometric redshift
errors, and the source clustering. For deep surveys, and with a
radial smoothing arising from the photometric redshifts, source
clustering can be safely ignored. We include the effects of
photometric errors, but ignore uncertainties in the photometric
redshift distribution. In terms of the observable photometric
redshift distribution of sources (all-sky), $\bar n(r) d^3\br = \bar
n_z(z_p) dz_p/4\pi$, we have
\begin{eqnarray}
\gamma(k,\bell) &\equiv& \sqrt{{1\over 8\pi^3}} \int dz_p\,
d^2\btheta \, \bar n_z(z_p)  \gamma(\br)\,kj_\ell(kr^0)
\\\nonumber & & \exp(-i\bell\cdot \btheta) W^0. \label{Gamma}
\end{eqnarray}

\subsection{Photometric redshift errors}\label{ss:photoz}
Photometric redshifts lead to a smoothing of the distribution in the
radial direction.  If we denote by $p(z_p|z)$ the probability of the
photometric redshift being $z_p$, given that the true redshift is
$z$, the mean of the expansion coefficients will be
\begin{eqnarray}
\gamma(k,\bell) &\equiv& \sqrt{{1\over 8\pi^3}} \int dz\,dz_p
d^2\btheta\, p(z_p|z)\,\bar n_z(z_p) \\\nonumber & &\gamma(\br)\,k
j_\ell(kr^0) \exp(-i\bell\cdot \btheta) W^0. \label{Gammap}
\end{eqnarray}
Note that $p(z_p|z)$ is arbitrary; it will generally have a
dispersion which depends on redshift, and can, if desired, include
broad wings to account for a small percentage of catastrophic
failures in the photometric redshift estimates. We assume a
Gaussian, with a $z$-dependent dispersion:
\begin{equation}
\label{Gauss}
p(z_p|z) = {1\over
\sqrt{2\pi}\sigma_z(z)}\exp\left[-(z_p-z+z_{\rm bias})^2\over
2\sigma_z^2(z)\right].
\end{equation}
$z_{\rm bias}$ is a possible bias in the photometric redshift
calibration, the effect of this on dark energy parameters is discussed
in Section \ref{Bias in the photometric redshifts}. Strictly the shear
is estimated at the actual radial coordinate of
the galaxy, which may differ from $r(z)$ because of peculiar
velocities.  We can safely ignore these, whose effect is small
compared with current photometric redshift errors.

\subsection{Relationship of \/ $\gamma(k,\bell)$ to cosmological
parameters}\label{ss:Gsec}
The lensing potential $\phi$ is related to the peculiar
gravitational potential $\Phi$ by a radial line-of-sight integral
(e.g. Bartelmann and Schneider, 2001):
\begin{equation}
\phi(\br) = {2\over c^2} \int_0^r dr' F_K(r,r') \Phi(\br').
\end{equation}
where $F_K(r,r')\equiv \left\{f_K(r-r')/
\left[f_K(r)f_K(r')\right]\right\}$, and $f_K(r)d\psi$ is the
dimensionless transverse comoving separation for points separated by
an angle $d\psi$.  The Robertson-Walker metric may be written $ds^2
= c^2 dt^2 - R^2(t)\left[dr^2 + f_K^2(r)d\psi^2\right]$, and
$f_K(r)$ takes the values $\sin r,\ r,\ \sinh r$ for curvature
values $k=1,0,-1$.  For a flat Universe $F_K(r,r')=(1/r'-1/r)$.

The peculiar gravitational potential is related to the overdensity
field $\delta(\br) \equiv \left[\rho(\br)-\bar\rho\right]/\bar\rho$
by the comoving Poisson's equation
\begin{equation}
\nabla^2\Phi = {3\Omega_m H_0^2\over 2 a(t)}\delta,
\label{PoissonReal}
\end{equation}
where $\Omega_m$ is the present-day matter density parameter, $H_0$
is the present Hubble constant and $a(t)=R(t)/R_0=1/(1+z)$ is the
scale factor.

Note that $\delta$ itself is not a homogeneous field, because it
evolves with time, and hence with distance from the observer through
the light travel time.  We get around the subtleties of this by
defining at each epoch a homogeneous field by referring all field
measurements to that time.  Thus we can, for example, define a power
spectrum which is time-dependent, and hence $r$-dependent.   This
may seem a little strange, since we have transformed from ${\bf r}$
space. The transforms of the homogeneous fields will be denoted by
$\delta(k,\bell;r)$ etc.

For high $\ell$, the transforms of $\Phi$ and $\delta$ (referred to
epoch $t$ or equivalently $r$) are related simply by
\begin{equation}
\Phi(k,\bell;r) = -{3\Omega_m H_0^2\over 2 k^2
a(t)}\delta(k,\bell;r).
\end{equation}
Inserting these definitions in the equation for $\gamma(k,\bell)$,we
find the relationship between $\gamma(k,\bell)$ and the transform of
$\delta$:
\begin{eqnarray}
\gamma(k,\bell) &=& -{3 \Omega_m H_0^2\over 2\pi^2 c^2}
\int_0^\infty dz dz_p p(z_p|z)
\\\nonumber & &  \!\!\!\!\!\!\!\!\!\!\! \int {d^2\btheta}\, k j_\ell(kr^0)W^0\,
\bar n_z(z_p)\exp(-i\bell.\btheta)
\\\nonumber & &  \!\!\!\!\!\!\!\!\!\!\!
\int_0^r dr' a^{-1}(r') \,
F_K(r,r')
\\\nonumber & & \!\!\!\!\!\!\!\!\!\!\!\int dk' {d^2\bell'\over (2\pi)^2}\,k'j_{\ell'}(k'r')
\delta(k',\bell';r')\frac{X_{\bell'}}{2} \exp(i\bell'.\btheta).
\end{eqnarray}
Integration over $\btheta$ gives $(2\pi)^2\delta^D(\bell-\bell')$,
so
\begin{eqnarray}
\label{GammaFinal}
\gamma(k,\bell) &=& -{3 X_\bell \Omega_m H_0^2\over 4\pi^2 c^2}
\int_0^\infty dz dz_p p(z_p|z) \bar n_z(z_p) \,
\\\nonumber & & \!\!\!\!\!\!\!\!\!\!\!\!\!\!\!\!\!\!\!\!\!\!\!
j_\ell(kr^0) W^0\int_0^r dr' F_K(r,r')(1+z')
\\\nonumber & & \!\!\!\!\!\!\!\!\!\!\!\!\!\!\!\!\!\!\!\!\!\!\!
\int dk' \,k'j_\ell(k'r')\, \delta(k',\bell;r'). 
\end{eqnarray}
This is a fundamental result of this paper.  It establishes the
connection between the (observable) 3D shear transform coefficients,
and the underlying matter density fluctuations, whose properties are
calculable from theory.

\subsection{Covariance matrix of $\gamma(k,\bell)$}\label{ss:Cov}
The signal part of the covariance matrix of the $\gamma(k,\bell)$ is
obtained from equation (\ref{GammaFinal}).  For the covariance of the
overdensity field coefficients, it is algebraically convenient to
use the geometric mean of the power spectra $P_\delta$, rather than
the power spectrum evaluated at epochs corresponding to $r$ or $r'$. 
Both of these could also be justified; note also that $P_\delta$
does not depend on $\bell$ (Castro, Heavens and Kitching, 2005).
\begin{eqnarray}
\langle \delta(k,\bell;r)\delta^*(k',\bell';r')\rangle &\simeq&
(2\pi)^2 \sqrt{P_\delta(k;r)P_\delta(k';r')}
\\\nonumber& & \delta^D(k-k') \delta^D(\bell-\bell').
\end{eqnarray}
The covariance matrix for the shear expansion coefficients is then
\begin{equation}
\langle\gamma(k,\bell)\gamma^*(k',\bell')\rangle_S =
Q_\ell\,(k,k')\,\delta^D(\bell-\bell')\label{Signal}
\end{equation}
where $Q_\ell\,(k,k')$ can be written as
\begin{equation}
Q_{\ell}(k,k') = {9 \Omega_m^2 H_0^4 X_{\bell}^2\over 4\pi^2 c^4} \int {d\tilde
k\over \tilde k^2}\, G_\ell(k,\tilde k) G_\ell(k',\tilde k)
\end{equation}
where
\begin{equation}
G_\ell(k,\tilde k) \equiv \int dz\,dz_p\, \bar n_z(z_p) W(z_p)
p(z_p|z) U_\ell(r,\tilde k) j_\ell(kr^0)
\end{equation}
and
\begin{equation}
U_\ell(r,k) \equiv \int_0^r d\tilde r \,{F_K(r,\tilde r)\over
a(\tilde r)} \sqrt{P_\delta(k; \tilde r)} \, j_\ell(k \tilde r).
\end{equation}
where $r=r(z)$ etc.  Equation (\ref{Signal}) is the second important
result of the paper.

\subsection{Shot noise}\label{ss:SN}
The shot noise can be calculated by making the usual assumption that
the galaxies are a Poisson sampling of an underlying smooth field
(see e.g. Peebles, 1980).  In practice we consider estimators of
the transforms of the individual components of the shear,
$\gamma_\alpha;\ \alpha=1,2$.  In the normal way for a point
process, these may be written as sums over small cells $c$, each of
which contains $n_c=0$ or 1 galaxy:
\begin{equation}
\hat\gamma_{\alpha}(k,\bell) \equiv \stp \sum_{\rm cells\ c} n_c
\gamma_{\alpha c}(\br_c)\,k j_\ell(k r^0_c) \exp(-i\bell\cdot
\btheta_c) W^0. \label{Gammahatcell}
\end{equation}
The variance of this involves a double sum over cells, and the
averaging over cells $c$ and $d$, $\langle n_c n_d \rangle$
contains shot noise terms when $c=d$, in which case $\langle
n_c^2\rangle = \langle n_c \rangle$, and the shot noise reduces to
a single sum, or an integral when we move back to a continuum
description.  Using the fact that the variance of the shear
estimate for a single galaxy is completely dominated by the
variance in the intrinsic ellipticity of the galaxy, $\sigma_e^2$,
rather than by lensing,
\begin{equation}
\langle \gamma_\alpha \gamma_\beta^* \rangle = {\sigma_e^2\over 2}
\delta^K_{\alpha\beta}
\end{equation}
where $\delta^K$ is a Kronecker delta function, and $\sigma_e\simeq
0.3$ (Brown et al., 2003), we find an expression for the shot noise as
\begin{eqnarray}
\label{Noise}
\langle
\hat\gamma_\alpha(k,\bell)\hat\gamma_\beta^*(k',\bell')\rangle_{SN}
& = & \sigma_e^2 \int dz\, \bar n_z(z) k
j_\ell(kr^0)\\\nonumber& & k'j_\ell(k'r^0)W^2(z) \,
\delta^K_{\alpha\beta} \, \delta^D(\bell-\bell').
\end{eqnarray}

\section{Estimation of cosmological parameters}
\label{s:params}
Cosmological parameters influence the shear transforms in a number
of ways: the matter power spectrum $P_\delta(k;t)$ is dependent on
$\Omega_m$, $h$ and the linear amplitude $\sigma_8$.  The linear
power spectrum is dependent on the growth rate, which also has some
sensitivity to the vacuum energy equation of state parameter $w(z)$.
$w$ also affects the $r(z)$ relation and hence the angular diameter
distance $f_K[r(z)]$. These parameters ($\{\Theta_\alpha\}$) may be
estimated from the data using likelihood methods.  Assuming uniform
priors for the parameters, the maximum a posteriori probability for
the parameters is given by the maximum likelihood solution.  We use
a Gaussian likelihood
\begin{equation}
2\ln L(\bg|\{\theta_\alpha\}) = {\rm constant}-\det(C)-\bg \cdot
C^{-1} \cdot \bg
\end{equation}
where $C=S+N$ is the covariance matrix, given by signal and noise
terms equations (\ref{Signal}) and (\ref{Noise}).  Note that
the average 
values of $\gamma(k,\bell)$ is zero, so the information on the
parameters comes from the dependence of the signal part of the
covariance matrix $C$. i.e. we adjust the parameters until the {\em
covariance } of the model matches that of the data.  This was the
approach of Heavens and Taylor (1995); Ballinger, Heavens and Taylor
(1995); Tadros et al. (1999); Percival et al. (2004) in analysis of
large-scale galaxy data. For many surveys, many useful modes of the
shear transform have contributions from wavenumbers where the power
spectrum is quite nonlinear.  The use of a Gaussian likelihood thus
needs to be justified by comparison with simulated data; this will
be explored in a later paper, and it is possible that a different
likelihood function may be necessary in the non-linear r\'egime.

\subsection{Expected errors on cosmological parameters - the Fisher
matrix}\label{ss:Fisher}
The expected errors on the parameters can be estimated with the
Fisher information matrix (Jungman et al., 1996; Tegmark, Taylor and
Heavens, 1997). This has the great 
advantage that different observational strategies can be analyzed
and this can be very valuable for experimental design. The Fisher
matrix gives the best errors to expect, and should be accurate if
the likelihood surface near the peak is adequately approximated
by a multivariate Gaussian.

The Fisher matrix is the expectation value of the second derivative
of the $\ln L$ with respect to the parameters $\Theta_\alpha$:
\begin{equation}
F_{\alpha\beta} = -\left\langle {\partial^2 \ln L\over \partial
\Theta_\alpha \partial\Theta_\beta}\right\rangle
\end{equation}
and the marginal error on parameter $\Theta_\alpha$ is
$\sqrt{(F^{-1})_{\alpha\alpha}}$.  If the means of the data are
fixed, the Fisher matrix can be calculated from the covariance
matrix and its derivatives (Tegmark, Taylor and Heavens, 1997) by
\begin{equation}
F_{\alpha\beta} = {1\over 2}{\rm
Trace}\left[C^{-1}C_{,\alpha}C^{-1}C_{,\beta}\right].
\label{Fisher1}
\end{equation}
For a square patch of sky, the Fourier transform leads to
uncorrelated modes, provided the modes are separated by $2\pi/L$
where $L$ is the side of the square in radians, and the Fisher
matrix is simply the sum of the Fisher matrices of each $\bell$
mode:
\begin{equation}
F_{\alpha\beta} = {1\over 2}\sum_\sbell {\rm
Trace}\left[(C^\sbell)^{-1}C^\sbell_{,\alpha}(C^\sbell)^{-1}C^\sbell_{,\beta}\right],
\label{Fisher2}
\end{equation}
where $C^\sbell$ is the covariance matrix for a given $\bell$ mode.
We compute $C^\sbell$ numerically from the signal and noise parts
equations (\ref{Signal}) and (\ref{Noise}), for given $\bar n_z(z)$,
photometric redshift error distribution, cosmology and survey area,
which governs the separation of uncorrelated $\bell$ modes.

\section{Survey Design Formalism}
\label{s:Errors}
In this Section we discuss the survey design factors.  We start by
detailing the 
assumptions of the survey design, and discuss some details of the
Fisher matrix calculation. Possible future weak lensing surveys and
their effectiveness are discussed in the Section \ref{Optimisation For
  a Wide-Field Lensing Survey}. 

\subsection{Survey parameters}
\label{ss:Survey}
In assigning survey parameters we follow the formalism detailed in
Taylor et al. (2006).
We assume that the redshift distribution for a typical
magnitude-limited survey is of the form
\begin{equation}
\bar n(z) \propto z^2 \exp\left[-\left({z\over
z_0}\right)^{1.5}\right]
\end{equation}
where $z_0 = z_m/1.412$, and $z_m$ is the median redshift of the
survey (e.g. Baugh and Efstathiou, 1993).
\begin{figure}
\psfig{file=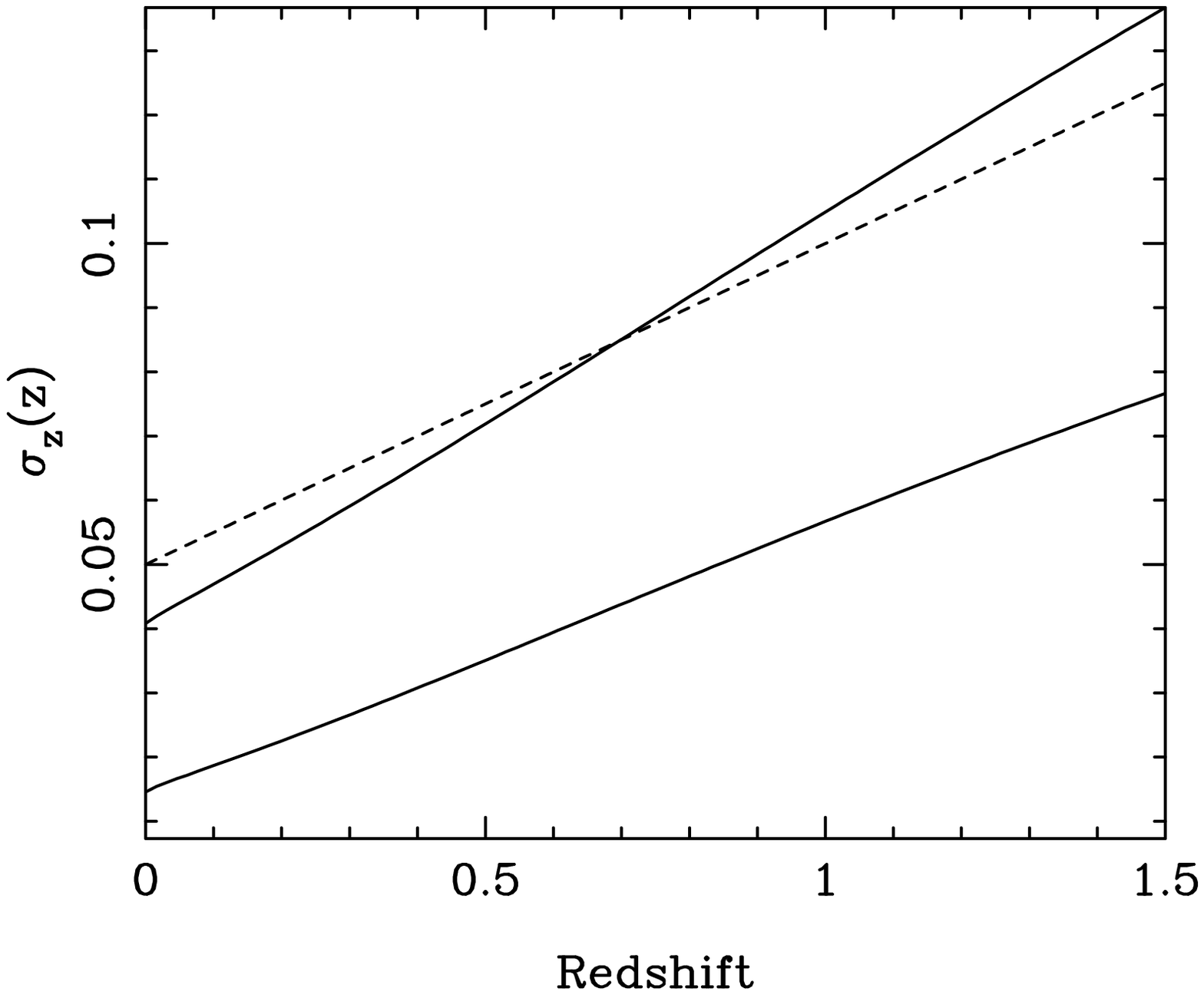, width=\columnwidth, angle=0}
\caption{The photometric errors as a function of
redshift $z$ for a 5-band (upper solid) and a 17-band (lower solid) optical
survey. The dashed line has $\sigma_z(z)=0.05(1+z)$ for comparison.}
\label{figsigma}
\end{figure}
The number density of useable sources with photometric redshift and shape
estimates is taken to scale as
\begin{equation}
n_0=30z_m^{3.4} \ \ {\rm per \ square \ arcminute}.
\end{equation}
This was estimated from the COMBO-17 survey. We take a maximum
redshift of $z_{\rm max}=1.5$ for ground based surveys. This 
is due to the difficulty of measuring galaxy shape, because of the
decrease in a galaxy's apparent size with increasing 
redshift, coupled with the seeing limit.

We assume that the photometric redshift errors are Gaussian, with a
dispersion given by $\sigma(z) = \sigma_0 (1+z) f(m,z)$, where $m$
is the apparent magnitude of the galaxy, the function $f(m,z)$ is
given in Taylor et al. (2006).  We integrate over a
Schechter function to get the average error as a function of $z$.
The error distribution is shown for a 5-band optical survey and a
17-band optical survey in Figure \ref{figsigma}.  The assumption of
Gaussianity of $p(z_p|z)$ can easily be relaxed: outliers can, for
example, be included, we investigate this in Section \ref{Additional
systematic effects for lensing}. We have extrapolated these formula
to $z=1.5$, though this extrapolation may be optimistic as
photometric redshift estimates can increases dramatically at
$z\approx 1$ if IR data is not available.

The variables which can be varied are the area $A$ and depth of the
survey ($z_m$), and the number of bands. These scale with the number
of nights observing $T$, the telescope diameter $D$ and the field-of-view
$F$ as (see Taylor et al., 2006)
\begin{equation}
\label{equaltime}
T\propto z_m^4 A D^{-2} F^{-1}.
\end{equation}
We consider as our default survey a $4$ metre telescope with
a $2$ square degree field-of-view which could observe
an area of $10$,$000$ square degrees to $z_m=0.7$ with $5$-bands in
$600$ nights of observing, this could be achievable with surveys such
as darkCAM (Taylor, 2005; conference proceedings of Probing the Dark
Universe with Subaru and Gemini) or the Dark Energy Survey (Wester, 2005).

\begin{figure}
\psfig{file=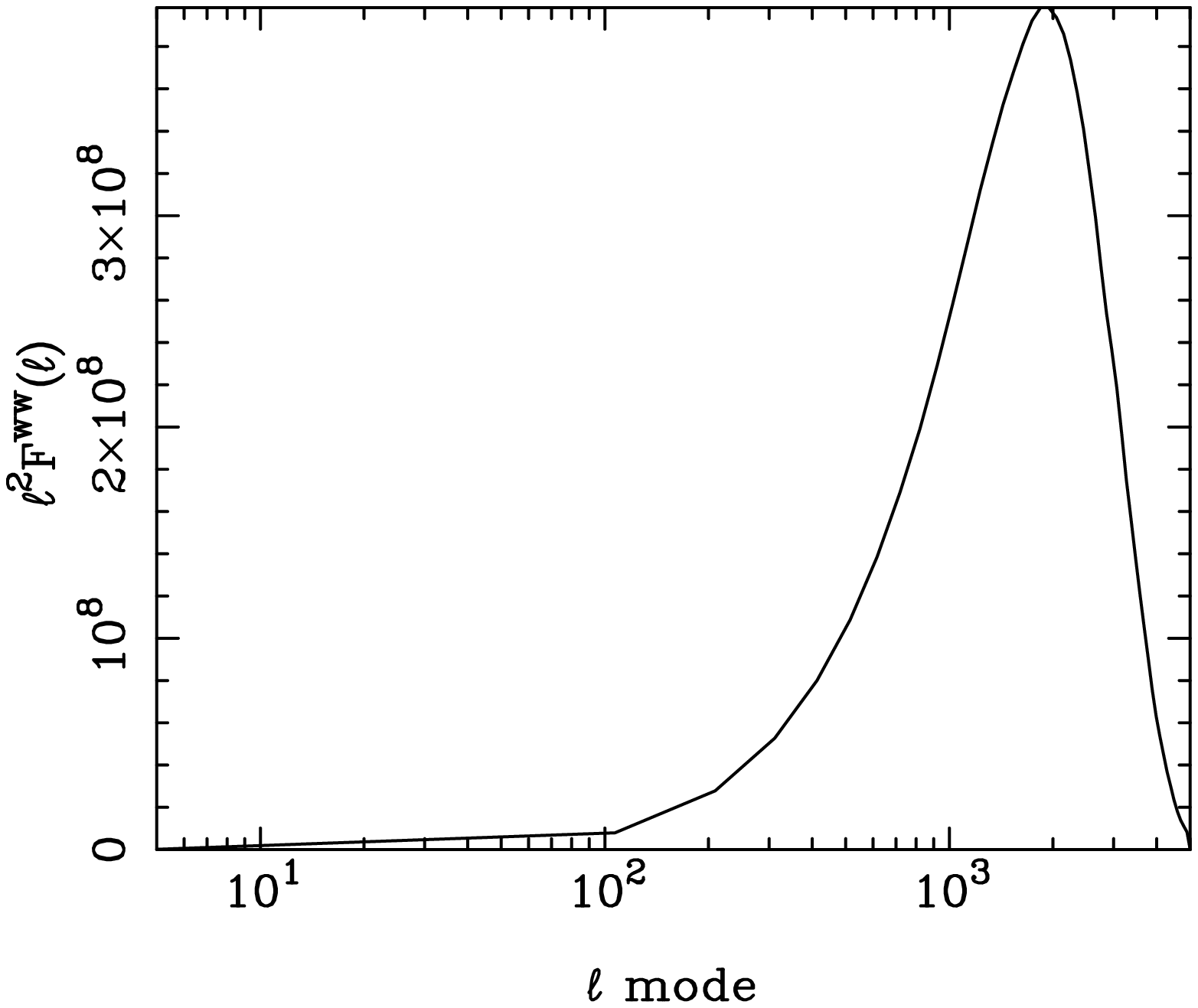, width=\columnwidth, angle=0}
\caption{The contributions to the Fisher matrix
element $F_{ww}$ from different ranges of $\ell$ for a 10,000 square
degree survey with $z_m=0.70$ and 5 bands. $w=w_0$ in this Figure for
clarity.} 
\label{FisherEll}
\end{figure}

We compute the nonlinear power spectrum using the fitting formulae
of Smith et al. (2003), based on linear growth rates given by
Linder and Jenkins (2003).  In order to avoid the high-$k$ r\'egime where
the formulae may be unreliable, or where baryonic effects might
alter the power spectrum ($k>10 h Mpc^{-1}$;
White, 2004; Zhan \& Knox, 2004), we do not analyse modes with
$k>1.5\,Mpc^{-1}$.  Note that the non-local nature of lensing does
mix modes to some degree, but these modes are sufficiently far from
the uncertain highly-nonlinear r\'egime that this is not a concern
(Castro, Heavens and Kitching, 2005). In any case the results are very
insensitive to the radial $k$ limit, since the photometric redshift
errors supress radial power at much lower $k$. We include angular
modes as  
small as each survey will allow, and analyse up to $\ell_{\rm
  max}=5000$. We note that the intracluster medium might affect
the power spectrum on the level of a few percent for $1000\ls\ell\ls
3000$ (Zhan \& Knox, 2004). These modes will still contain useful
information, but a more detailed analysis might be nessacary when the
method is applied. To help asses the extent of any modification to the
expected accuracy, we
also quote results for a more conservative limit of $\ell_{\rm
  max}=2000$, this increases the predicted marginalised errors by
approximately $0.01$. The flat-sky approximation will 
break down for the low-$\ell$ modes, but there is little power there
in any case (Figure \ref{FisherEll}).

We allow for a Universe with the following parameters: $\Omega_m$,
$\Omega_v$, $h$, $\sigma_8$, $\Omega_b$, $w_0$, $w_a$, $n_s$ and $\alpha_n$.
$\sigma_8$ represents the amplitude of the perturbations, $n_s$ the
scalar spectral index and its running $\alpha_n$, and we parameterize
$w(a)$ by equation 
(\ref{DEeqstate}). Note that our
assumption of this form is not critical; theoretical models with
arbitrary $w(a)$ can be analyzed. We choose this form to investigate
the sensitivity of our results on $w_0$ and $w_a$ to
time-dependence.

\section{Parameter Forecasts and Optimization For a Wide-Field Lensing Survey}
\label{Optimisation For a Wide-Field Lensing Survey} 
Having
introduced the method and the survey design formalism this Section
will investigating optimizing a weak lensing survey so that the
marginal errors on the dark energy parameters can be minimized. We
will explore the variation in the marginal error on $w_0$ with changes
in the median depth, varying the area to
preserve the total observation time, and the redshift error.

\subsection{Combining with other dark energy experiments}
\label{Combining with other dark energy experiments} 
There are a
number of alternative ways to place constraints on the dark energy
equation of state parameters. Prominent among these are the CMB,
Baryon Acoustic Oscillations (BAO) in the galaxy power spectrum and
the supernova Type Ia (SNIa) Hubble diagram. Each of these methods
can place constraints on the dark energy equation of state although
they all suffer from degeneracies between $w_0$ and $w_a$. In
Section \ref{Parameter Forecasts} we show how by combining the
different methods the parameter degeneracies can be lifted allowing
for improved accuracies on $w_0$ and $w_a$.

Using the Fisher matrix methods outlined in detail in Taylor et
al. (2006) we calculate predicted Fisher matrices and 
parameter constraints for the following surveys. In all the Fisher
matrix calculations we use an 11 parameter cosmological set
($\Omega_m$, $\Omega_v$, $h$, $\sigma_8$, $\Omega_b$, $w_0$,
$w_a$, $n_s$, $\tau$, $\alpha_n$, $r=T/S$), with default values
(0.27, 0.73, 0.71, 0.8, 0.04, -1.0,0.0, 1.0, 0.09, 0.0, 0.01). For a
summary of the main assumptions that went into each of the Fisher
matrix calculations see Table \ref{planckparam}.
\begin{table}
\begin{center}
\begin{tabular}{|l|c|c|c}
 {\bf Lensing Survey}&&&\\
 \hline
 Area/sq degrees&$z_m$&$z_{\rm max}$&$N_{\rm Bands}$\\
 \hline
 $10$,$000$&$0.70$&$1.5$&5\\
 \hline
 \\
 \\
 \hline
 {\bf Planck}&&&\\
 \hline
 Band/GHz & $\theta_{{\rm beam}}$ &
 $\sigma_T$/$10^{-6}$ & $\sigma_P$/$10^{-6}$\\ 
 \hline
 $44$& $23'$&$2.4$&$3.4$\\
 $70$& $14'$&$3.6$&$5.1$\\
 $143$&$8.0'$&$2.0$&$3.7$\\
 $217$&$5.5'$&$4.3$&$8.9$\\
 \hline
 \\
 \\
 \hline
 {\bf WFMOS}&&&\\
 \hline
 Area/sq degrees&$z_{\rm bin}$&$k_{\rm max}$/$hMpc^{-1}$&Bias\\
 \hline
 $2000$&$1.0$&$0.15$&$1.25$\\
 $300$&$3.0$&$0.15$&$1.25$\\
 \hline
 \\
 \\
 \hline
 {\bf SNAP}&&&\\
 \hline
 $z_{\rm max}$&$N_{\rm bin}$&$N_{SNIa}$&$\sigma_m$\\
 \hline
 $1.5$&$17$&$2000$&$0.15$\\
 \hline
\end{tabular}
\caption{The main default values parameterising the Lensing, CMB,
  BAO and SNIa experiments considered in this paper.}
\label{planckparam}
\end{center}
\end{table}

We consider a $4$-year WMAP experiment and a $14$-month Planck
experiment (Lamarre et al., 2003) which will be 
contemporary with the type of wide field photometric surveys being
considered for 3D weak lensing. We computed the CMB covariance
matrices by using CMBFAST (Seljak and Zaldarriaga, 1996), the survey parameters
were taken from Hu (2002). We include polarisation but do not
marginalize over the calibration error. Also we do not consider the
Integrated Sachs-Wolfe (ISW) effect directly via cross-correlating
with galaxy surveys, although this will also provide an interesting
dark energy constraint.

For a BAO experiment we consider the Wide Field Multi
Object Spectrograph (WFMOS; Bassett et al., 2005) following the
methodology outlined in Seo \& Eisenstein (2002), Blake \& Glazebrook
(2003) and Wang (2006). 

We use the method outlined in Ishak (2005) and Y\`eche et al. (2006)
to calculate a SNAP (Aldering, 2005) SNIa Fisher matrix. The effective
magnitude uncertainty takes into account luminosity evolution,
gravitational lensing, dust and the effect of peculiar velocities.

\subsection{Combining with the CMB alone}
\label{Combining with the CMB} To help to lift degeneracies between
cosmological parameters, and to retain realism in our predictions
for a wide field photometric survey we consider a $14$-month Planck
CMB experiment. This CMB experiment will place constraints on $w_0$
and $w_a$, mainly through the large scale ISW effect, although there
is a strong degeneracy between the dark energy parameters. Combining
with 3D weak lensing helps to lift this degeneracy.

\begin{figure*}
\psfig{file=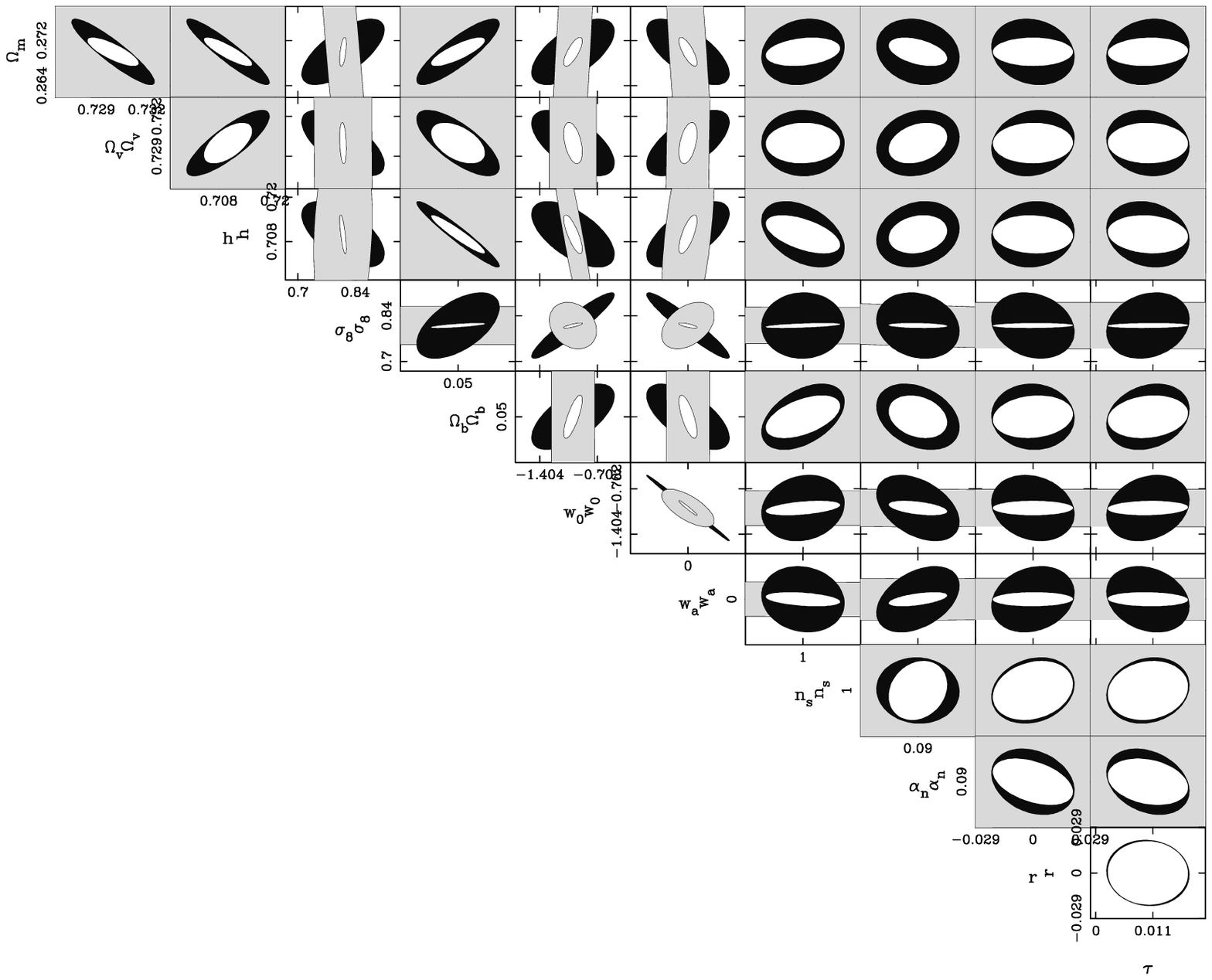, width=2\columnwidth, angle=0}
\caption{Expected marginal errors on cosmological parameters from
Planck (dark), 3D weak lensing survey (light gray)
and the combination (central, white).  The survey covers $10,000$ square
degrees to a median depth of $0.7$ in $5$ bands. Ellipses show the
$1$-$\sigma$ errors for two parameters (68.3\% confidence regions),
marginalized over all other parameters.}
\label{figdarkCAM}
\end{figure*}

For the default survey, we show in Figure \ref{figdarkCAM} the
Fisher matrix elements marginalized over all other parameters. The
darkest areas show a 14-month Planck
prior. The light gray ellipses show the
two-parameter,
$1$-$\sigma$ errors for the parameters plotted, and the white (central)
ellipses show the combination.  The marginal errors on $w_0$ and $w_a$
are $\Delta w_0=0.108$ and $\Delta w_a=0.395$ respectively, a factor
of $5$ improvement over the $14$-month Planck constraints alone which
could constrain $w_0$ and $w_a$ to $\Delta w_0=0.502$ and $\Delta
w_a=1.86$.
\begin{figure}
\psfig{figure=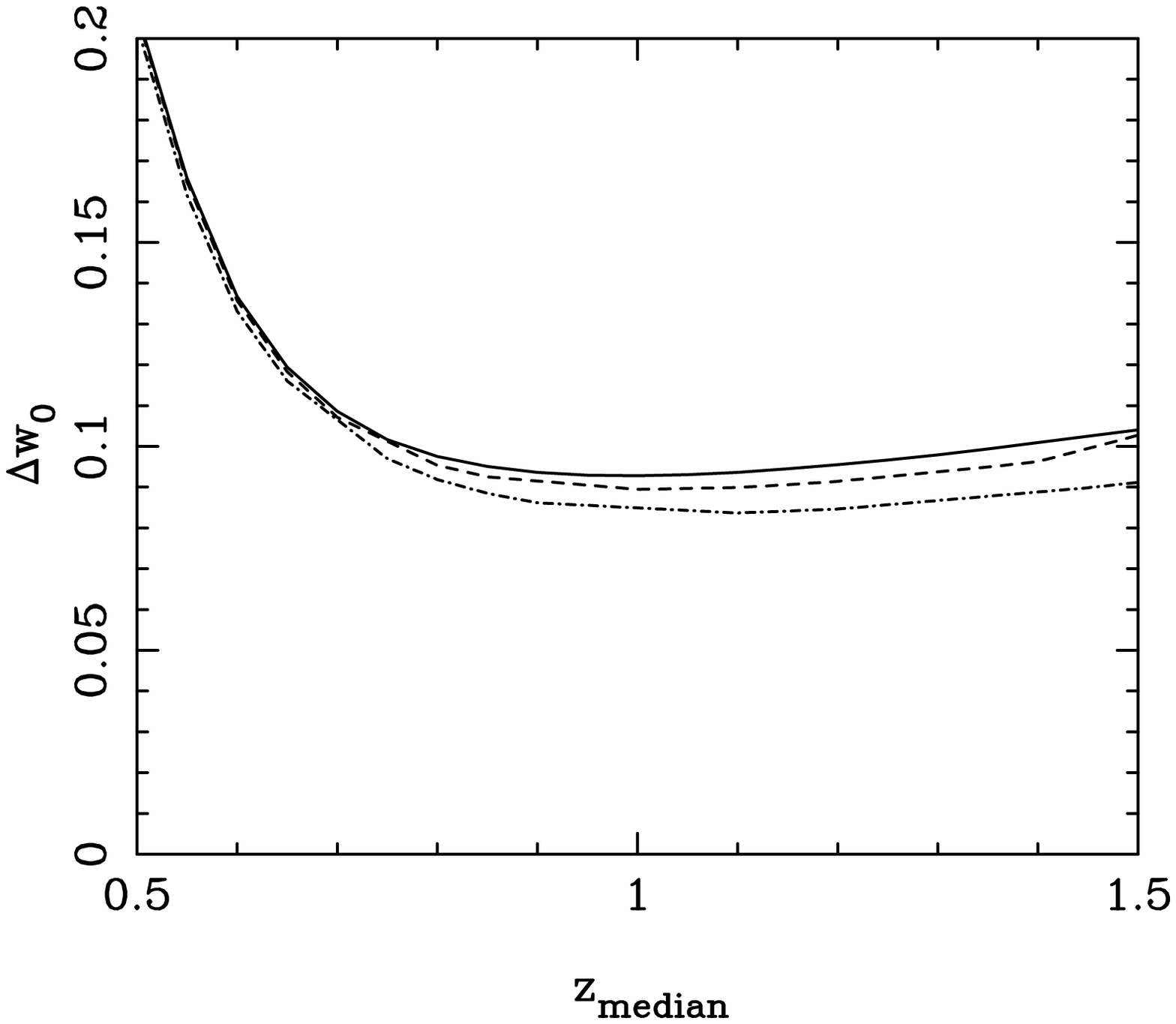, width=\columnwidth,angle=0}
\caption{The variation in the marginal error on $w_0$ as the median
redshift of the survey varies for a $600$ night survey on a $4$ metre class
telescope, including a 14-month Planck prior. Note
we assume shapes are note measurable beyond $z_{\rm max}=1.5$. The
solid line is for  
a $5$ band survey, the dashed line for a $9$ band survey and the
dot-dashed line for a $17$ band survey.}
\label{zmedian}
\end{figure}

3D weak lensing improves constraints on all the CMB cosmological
parameters, in particular $\sigma_8$ whose constraint is improved by
a factor of $14$. It is already well known that weak lensing can
tightly constrain the ($\sigma_8$, $\Omega_m$) plane, using standard
cosmic shear techniques (see Brown et al., 2003; Semboloni et al., 2006),
3D weak lensing constrains
$\sigma_8$ in the same way by measuring the overall normalisation of
the matter power spectrum. 3D weak lensing provides indirect (and slight)
improvements on the constraints for the tensor to scalar ratio $r$
and the optical depth to last scattering $\tau$ through the
intersection of the 3D weak lensing multi parameter ellipsoid with
the CMB's multi parameter constraint. Table \ref{planckimprov} shows
the $14$-month Planck constraints and the new combined constraints
once 3D weak lensing is included. Note that results are presented for
Universes which are not necessarily flat. In non-flat geometries the
spherical Bessel functions $j_{\ell}(kr)$ should be replaced by
ultra-spherical Bessel 
functions $\Phi^{\ell}_{\beta}(y)$. For the case considered here
$\ell\gg 1$ and 
$k\gg$ (curvature scale)$^{-1}$ then $\Phi^{\ell}_{\beta}(y)\rightarrow
j_{\ell}(kr)$ (Abbott and Schaefer, 1986; Zaladarriaga and Seljak,
2000). The expansion used is not  
ideal for non-flat Universes but should be an adequate approximation
given current constraints on flatness. 
\begin{table}
\begin{center}
\begin{tabular}{|l|c|c|c|}
\hline
Parameter&Planck only&Lensing only&Combined\\
\hline
$\Omega_m$&$0.0058$&$0.0500$ &$0.0025$\\
\hline
$\Omega_v$&$0.0024$&$0.0795$ &$0.0015$\\
\hline
h&$0.0088$&$0.0321$&$0.0051$\\
\hline
$\sigma_8$&$0.1002$&$0.0705$&$0.0073$\\
\hline
$\Omega_b$&$0.0011$&$0.3707$ &$0.0007$\\
\hline
$w_0$&$0.5015$&$0.2843$&$0.1086$\\
\hline
$w_a$&$1.8618$&$1.1792$&$0.3947$\\
\hline
$n_s$&$0.0034$&$0.3852$&$0.0031$\\
 \hline
$\alpha_n$&$0.0062$&$0.0576$&$0.0045$\\
 \hline
$\tau$&$0.0079$& &$0.0077$\\
 \hline
$r $&$0.0208$& & $0.0203$\\
 \hline
\end{tabular}
\caption{Improvements on CMB Planck one parameter 1-$\sigma$,
  constraints by
  adding 3D weak lensing from a 10,000 square degree lensing
  survey to a median depth of $z_m=0.7$.}
\label{planckimprov}
\end{center}
\end{table}

\subsection{Survey Optimization}
\label{Survey Optimisation}
For a given observing time, there will
be an optimum depth of survey to minimize the statistical error on
$w_0$ (or $w_a$). A very wide, shallow survey will yield poor
cosmological constraints since the lensing signal is so small,
whereas a very deep survey will also yield poor cosmological
constraints because very little area can be covered, and the cosmic
variance will be large. In addition, the distant galaxies will have
shapes which are difficult to measure at high redshift. Here we
explore the optimum median redshift using equation (\ref{equaltime})
keeping the time fixed, so that the area of the survey scales with
median redshift as $A\propto z_m^{-4}$. The results are for $600$
nights on a 4m survey telescope with a 2 square degree
field-of-view, where $z_m=0.70$ corresponds to $A=10,000$ square
degrees. The results are shown in Figure \ref{zmedian}.

The optimal median redshift for a $5$ band optical survey is $z_m=1.0$
with an area of $A=2400$
square degrees. The error is poor below $z_m\sim 1.0$ because the lensing
signal becomes weaker, as the number of lensed background galaxies at
high redshift decreases. For median redshifts above $z_m\sim 1.0$ the
error is also poor as shot noise begins to dominate and the areal coverage
becomes too small. We also
investigate the effect of using $9$ and $17$
optical bands, keeping the median redshift and area fixed. We find
that the marginal error does not substantially decrease. This is due
to the combination of the lensing and CMB, the intersection of the
ellipses in parameter space, remaining similar even though the lensing
marginal errors on their own decrease with increasing bands. The extra
bands might well be useful in the identification of outliers which we
discuss in Section \ref{Additional systematic effects for lensing}.

The optimal median redshift increases with the number of optical bands,
a $17$ band optical survey has an optimal median redshift of $z_m=1.1$
with an area of $A=1640$ square degrees. Higher-$k$ modes are
accessible due to the reduced damping effect of the photometric
redshift error. In order to utilise these modes, the shot noise needs
to be reduced, so the optimisation favours a slightly deeper survey
with higher number density. This effect increases as the redshift
uncertainty decreases. 

Taylor et al. (2006) find that the optimal survey design
for the geometric ratio method is $z_m=0.70$ and $A=10,000$ square
degrees for a $5$ band photometric redshift survey on a $4$ metre
telescope with a $2$ square degree field-of-view. So that the
results presented here can be directly comparable with their results
we shall adopt a fiducial survey design of  $z_m=0.70$ and
$A=10,000$ square degrees in $5$ bands from hereon; although it
should be noted that given $600$ nights on such an instrument that
an optimal survey design could improve the marginal errors on $w_0$
and $w_a$ using this method.

\subsection{Weighting the data}
\label{Weighting the data} 
\begin{figure}
\psfig{file=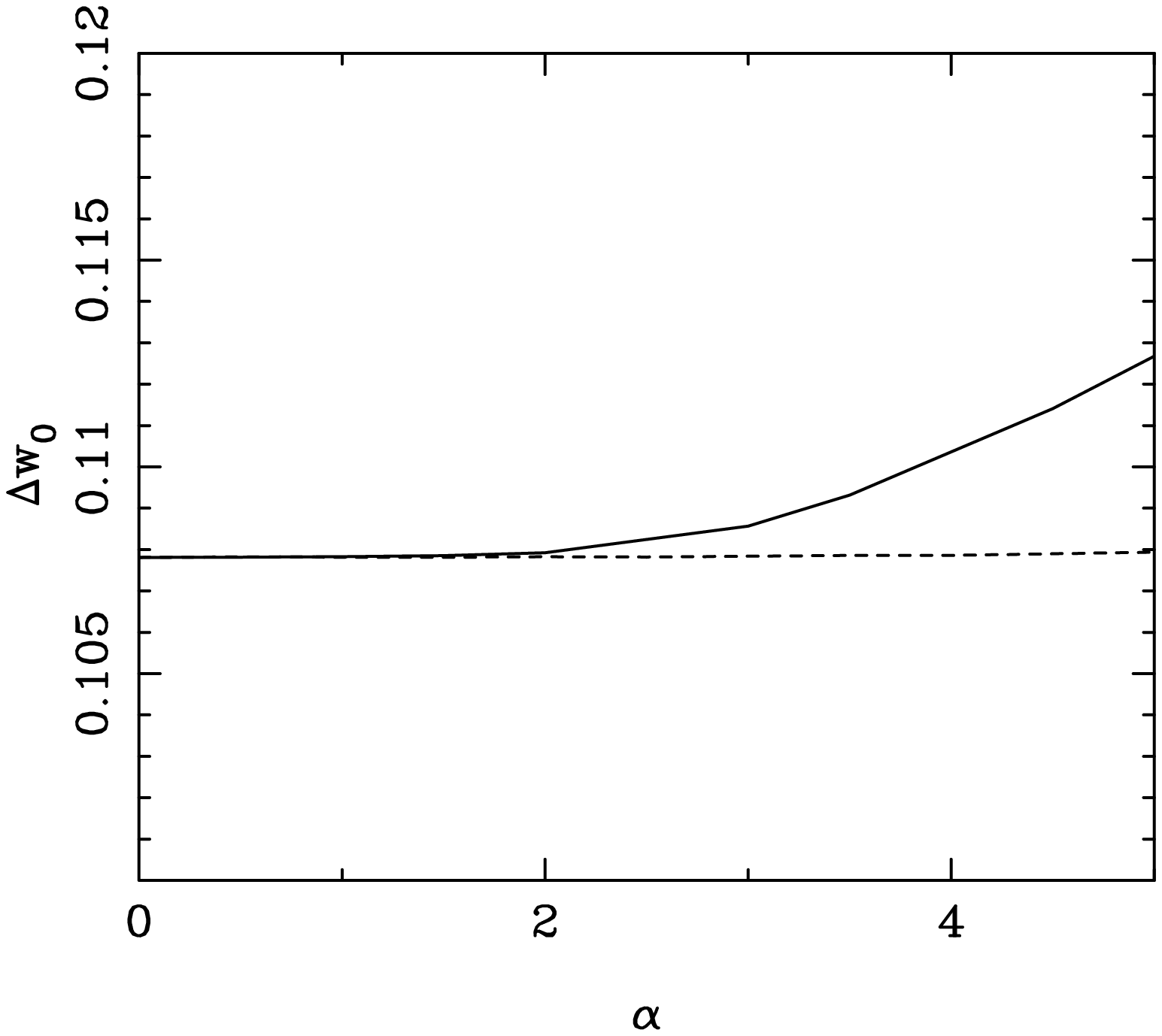, width=\columnwidth, angle=0}
\caption{The variation in the lensing-only marginal error on $w_0$,
as the weighting scheme is varied. The solid line is for $W(z) =
z^\alpha$, the dashed line is for $W(z) = (1+z)^\alpha$; $\alpha$ is
varied.} \label{weighting}
\end{figure}

We might expect that the cosmological
parameter constraints could be improved if the more distant galaxies
are given higher weight. We
have not attempted a formal optimization, but show some results of
experimental weighting schemes.  We consider two weighting schemes
$W(z)=z^\alpha$ and $W(z)=(1+z)^\alpha$. Figure \ref{weighting}
shows how the error on $w_0$ changes as the weighting scheme is
changed. The figure shows the lensing-only marginal combined with a
14-month Planck prior on all parameters.
We show that the marginal error
on $w_0$ is fairly insensitive to the weighting scheme
employed. Furthermore using equal weighting is in fact the optimal
strategy for a weighting functional form of this kind. This shows that the
increase in the shot noise through the weighting of high redshift galaxies
counteracts any improvement in the lensing signal, used to constrain $w_0$.

\subsection{Optical and Infrared surveys}
\label{Optical and Infrared surveys} By combining a $5$ band optical
survey with, for example, a $4$ band infrared survey the photometric
redshift accuracy can decrease. Strategies such as this have the
potential to be employed on future wide field surveys in an effort
to improve cosmological parameter constraint. Here we parameterize
the redshift error using
\begin{equation}
\sigma_z(z)=\sigma_0(1+z)
\end{equation}
where a $5$ band optical survey can be approximately represented,
see Figure \ref{figsigma}, by $\sigma_0=0.05$. $\sigma_0=0.01$
corresponds to a $9$ band survey comprising of $5$ optical and $4$
infrared bands (Wolf, private communication). Note the distinction
between this and the $9$ band optical 
(no infrared) survey considered in Section \ref{Survey Optimisation}.

\begin{figure}
\psfig{file=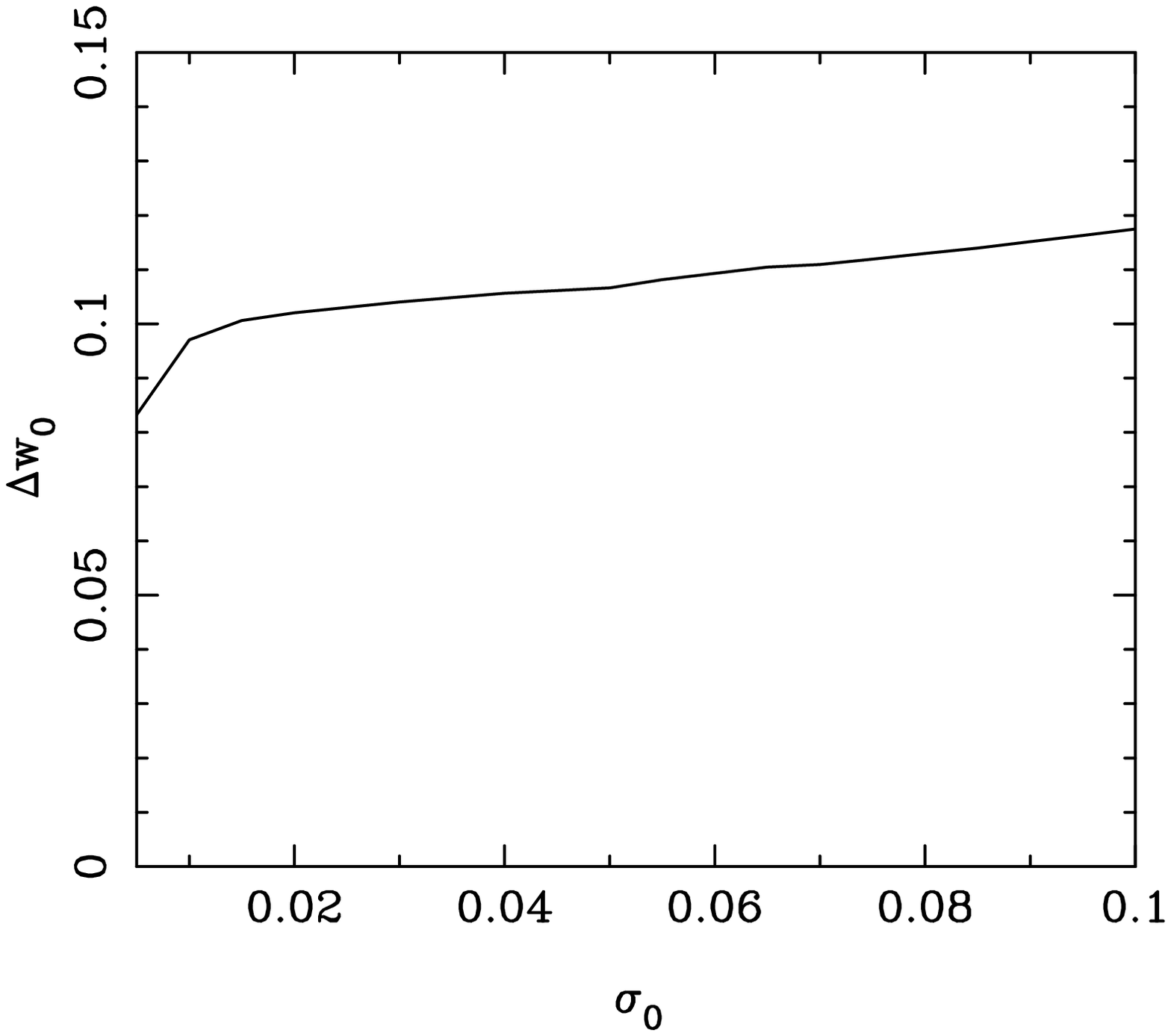,width=\columnwidth,angle=0}
\caption{Marginal error on $w_0$ for different photometric redshift
errors parameterized by $\sigma_z(z)=\sigma_0(1+z)$. These results
include a $14$-month Planck prior.} \label{optimalsigma}
\end{figure}

Figure \ref{optimalsigma} shows how the marginal error on $w_0$ varies
with $\sigma_0$, here we fix the survey design to be $z_m=0.7$ and
$A=10,000$ square degrees.
We find that the marginal error on $w_0$ varies slowly between
$0.01<\sigma_0<0.1$ and improves rapidly for $\sigma_0<0.01$. This
turn-over corresponds to the point where the lensing pivot redshift
error, see Section \ref{Pivot redshifts}, becomes comparable to the
CMB pivot redshift error. So, the marginal error of the combined
constraint improves at a faster rate (as $\sigma_0$ decreases)
after this point since the lensing constraint is improving $w_0$ and
$w_a$ and lifting the CMB degeneracies further.
\begin{figure}
\psfig{file=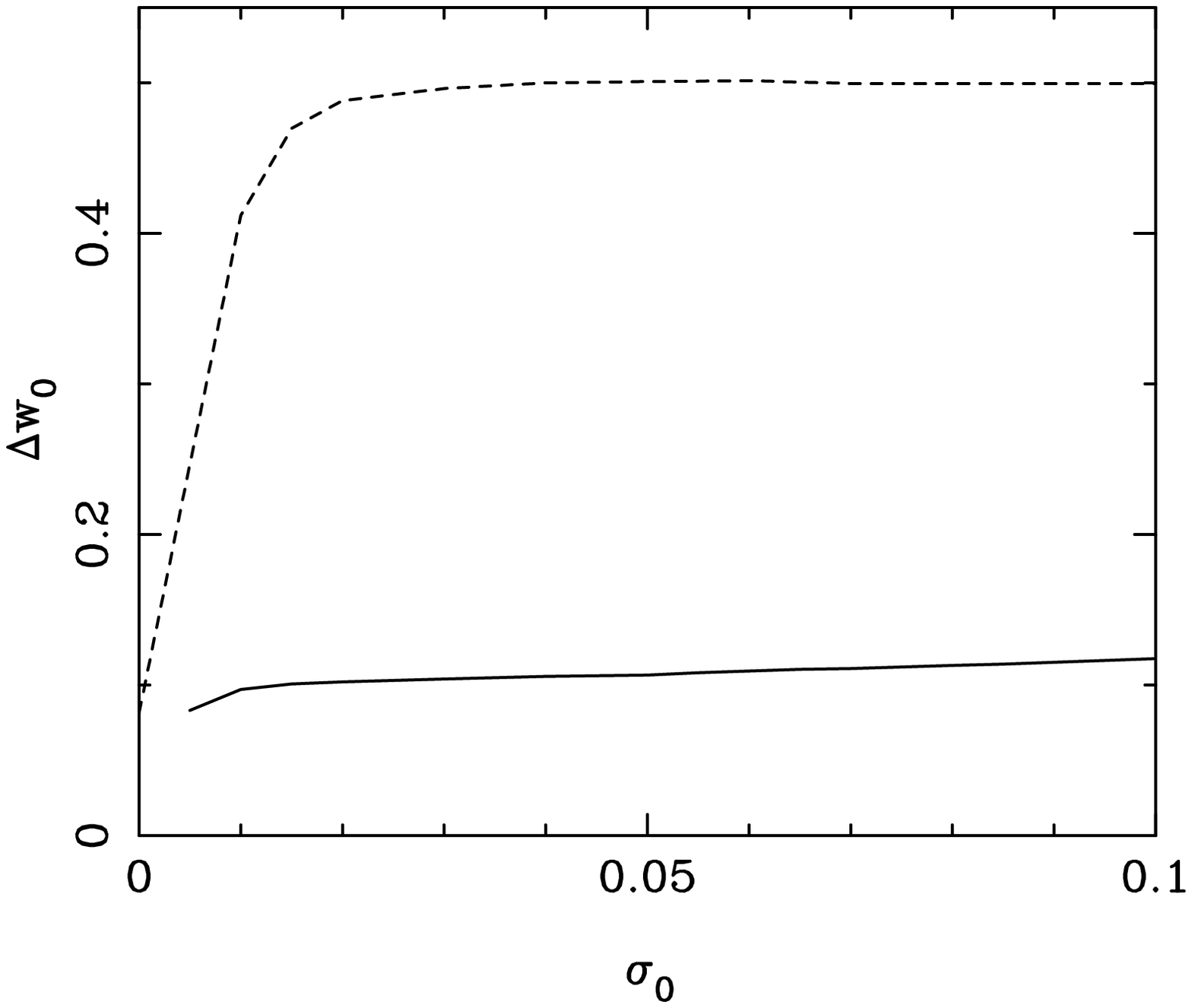,width=\columnwidth,angle=0}
\caption{Marginal error on $w_0$ for different photometric redshift
errors parameterized by $\sigma_z(z)=\sigma_0(1+z)$ for lensing
(solid line) and a BAO experiment (dashed line) from survey of
$z_m=0.70$ and 
$A=10,000$ square degrees. These results include a $14$-month Planck
prior.} \label{optimalsigmaBAO}
\end{figure}
Figure \ref{optimalsigmaBAO} shows the marginal error from both
lensing and BAO using the photometric redshifts from our fiducial
survey. The treatment of a BAO experiment in a photometric redshift
survey is discussed in Taylor et al. (2006). We find
that BAO constraints do not significantly improve the $14$-month
Planck prior for $\sigma_0>0.02$, and that for a $5$ band survey
lensing provides much tighter constraints on the dark energy
parameters. The dark energy constraints from lensing are less
effected by the photometric damping of the radial wavenumber due to
poor photometric errors than the BAO since lensing also gains dark
energy information from geometric factor via the lensing effect. The
BAO methodology on the other hand relies on a good measurement of
the power spectrum which is restricted to low wavenumbers due to the
photometric redshift errors. At low $\sigma_0$ the two methods
provide complimentary constraints on $w_0$.

\subsubsection{Bias in the photometric redshifts}
\label{Bias in the photometric redshifts}
As well as investigating the effect of varying the absolute values of
the photometric errors, we would also like to investigate how a bias
in the photometric redshift calibration effects the dark energy
parameter estimation. Knox, Scoccimarro and Dodelson (1998), Kim et
al. (2004) and Taylor et al. (2006) show how the bias $\delta\psi_j$
on a fixed model 
parameter $\psi_j$ is related to the marginal error on a (cosmological)
parameter $\delta\theta_i$ by
\begin{equation}
\delta\theta_i=-[F^{\theta\theta}]_{ik}^{-1}F_{kj}^{\theta\psi}\delta\psi_j.
\end{equation}
$F^{\theta\theta}$ is the cosmological parameter Fisher matrix and
$F^{\theta\psi}$ is a pseudo-Fisher matrix between the measured and
assumed parameters, when considering one parameter this is a column
matrix.

We assume that there is some bias $z_{\rm bias}$ in the mean of the
photometric redshifts in a given survey (see equation (\ref{Gauss}))
due to poor calibration of photometric redshifts with a
spectroscopic training set. Marginalizing over all other parameters
we find that
\begin{equation}
\delta w_0=-C_{\rm bias}\delta z_{\rm bias},
\end{equation}
where $C_{\rm bias}$ is some constant. Following the arguments in
Taylor et al. (2006) the number of galaxies requiring
spectroscopic redshifts is
\begin{equation}
N_{\rm spec}=\left[\frac{C_{\rm bias}\sigma(z)}{\delta w_0}\right]^2,
\end{equation}
where the bias on $w_0$ is half the error $\delta w_0=0.5\Delta
w_0$. We have found that $C_{\rm bias}\approx 1.2$ for 3D weak
lensing. If $\sigma(z)\approx 0.1$ and we require $\Delta w_0\approx
0.01$ the number of spectroscopic redshift required is $N_{\rm
  spec}\approx 6\times 10^{2}$. This number is easily
achievable using the current generation of spectrometers. The number
of required spectroscopic redshifts is 
significantly smaller than the large number required for the
geometric ratio test, for which $C_{\rm bias}\approx 9.0$ (e.g. Taylor
et al., 2006) and tomographic methods (e.g. Hu and Jain, 2004), we
attribute this difference to the binning procedure required in
these methods. In binning the data any offset in the redshift
estimation of a galaxy will create a discrepancy between the
estimated and actual number of galaxies in a bin, and any derived
quantities gained from them, for example the tangential shear behind
a cluster. In this analysis any systematic offset in the galaxy
population does not affect any derived quantities in this way but
rather the whole shear field is offset in redshift, and galaxies are
simply given a slightly increased weighting via $j_{\ell}$. Note that
we consider only a single bias parameter, rather than the more complex
behaviour allowed in Ma et al. (2005). However, it is the same model
as used in our companion paper (Taylor et al., 2006) which shows more
sensitivity to $z_{\rm bias}$.

Following the procedure outlined in Taylor et al. (2006) we also
investigated the effect of an offset in the variance of the 
photometric redshift errors
$\sigma_z(z)\rightarrow\sqrt{(\sigma_z^2(z)+\Delta\sigma_z^2(z))}$.
We find that this effect is negligible for this analysis, so that
the total bias due to photometric redshift errors is only dependent
on the bias in the offset of the mean. We will explore fully
marginalizing over nuisance parameters in a full Fisher analysis
elsewhere. 

\subsubsection{The effect of outliers}
\label{Additional systematic effects for lensing}
\begin{figure}
\psfig{file=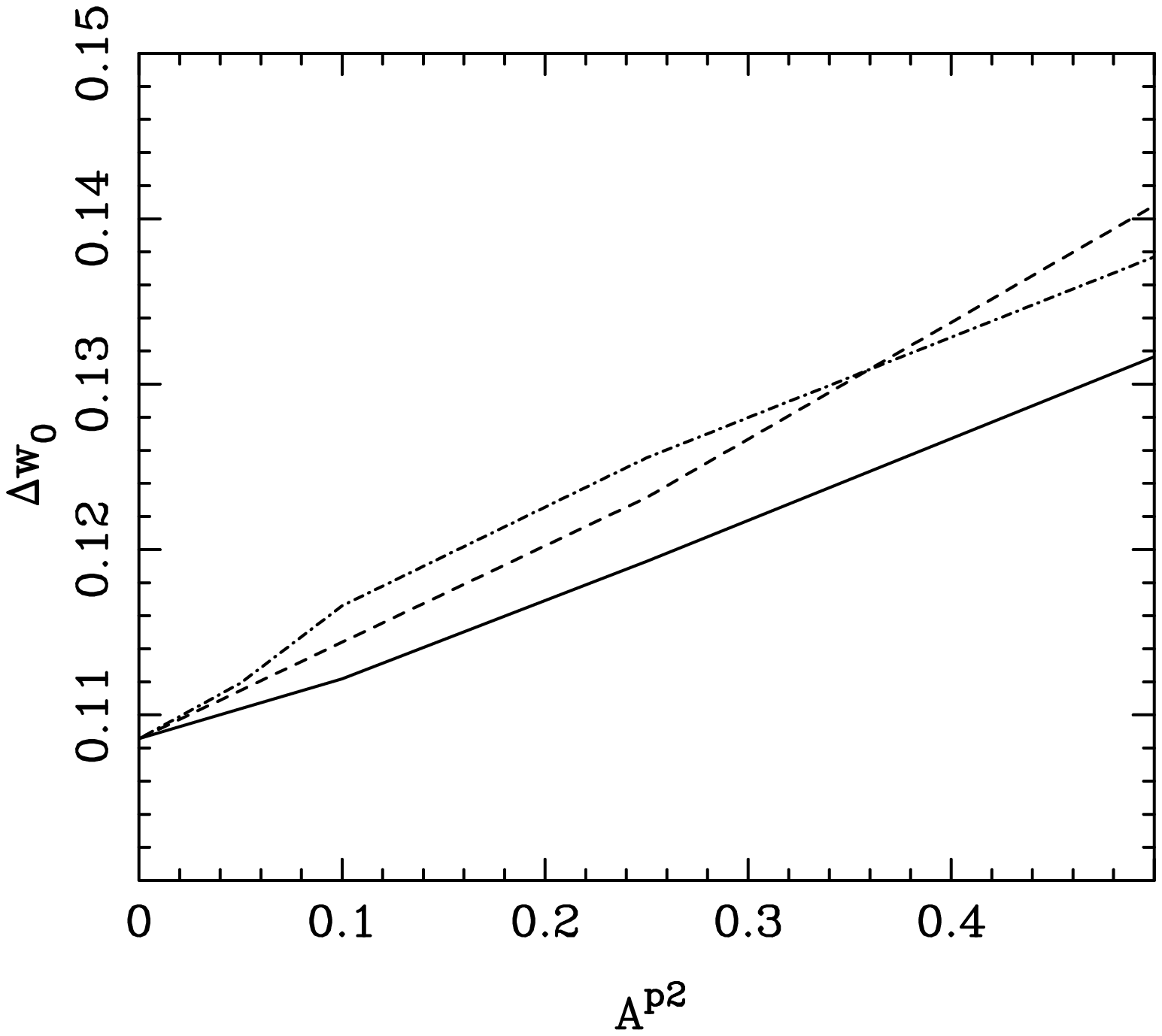,width=\columnwidth,angle=0}
\caption{The effect of outliers with $\sigma^{p2}_z(z)=0.5$ on the
  marginal error in $w_0$ as a 
  function of the proportion of outliers in the survey $A^{p2}$. The
  survey is a $5$ band survey of 10,000 square degrees to a median
  redshift of $z_m=0.7$, with a $14$-month Planck prior. The solid
  line shows the effect of treating the outliers as a separate
  population; the dashed line shows the effect of discarding the
  outliers; the dot-dashed line shows the effect of incorporating the
  outliers into a single galaxy population.}
\label{outliers}
\end{figure}

In every photometric redshift survey there will be some galaxies in
the sample for which an accurate photometric redshift cannot be
assigned. To investigate the effect of these outliers on the dark
energy parameter estimation we consider two galaxy populations, one
with the original photometric redshift errors, see Section
\ref{ss:Survey}, and a second population with
$\sigma_z^{p2}(z)=0.5$. We show the results in Figure \ref{outliers},
$A^{p2}$ is the proportion of the total galaxy population with
$\sigma_z^{p2}=0.5$. We have considered three ways in which the outlying
population could be dealt with, and how the effect of each of these
methods varies with the proportion of the total population of outlying
galaxies.  

A population of outliers can either be discarded from the analysis
completely or used in some way. The dashed line in Figure
\ref{outliers} shows the effect of discarding the sample, so that the
surface number density of galaxies is decreased by $n_0\rightarrow
n_0(1-A^{p2})$, but the photometric redshift error remains the
same. To use the outliers either they can be treated as a
separate population (solid line in Figure \ref{outliers}) or can be
incorporated into a single population (dot-dashed line in Figure
\ref{outliers}) in which case the overall 
photometric redshift is degraded to
$\sigma_z(z)\rightarrow\sqrt{(1-A^{p2})[\sigma_z(z)]^2+A^{p2}[\sigma_z^{p2}(z)]^2}$
(see Blake \& Bridle, 2005) where $\sigma_z(z)$ is the original
photometric redshift error.
The effect of having outliers in the sample increases the marginal
error on $w_0$ regardless of how they are treated, though the method
is relatively insensitive to this effect. As expected using
the outlying galaxies, and treating them as a 
separate population, increases the marginal error less than discarding
the galaxies completely. By incorporating the galaxies into a single
population the redshift error is degraded to such a degree that for a
low proportion of outliers it is optimal to discard them, note the
signal-to-noise for 3D weak lensing is proportional to $n_0$. For a high
proportion it is optimal to include them somehow either into a single
population or as two separate populations. 

\subsection{Scaling to other Surveys}
\label{Scaling to other Surveys}
Using the results presented in Sections \ref{Survey Optimisation} and
\ref{Optical and Infrared surveys} we can 
provide scaling relations in a similar way to Taylor et al. (2006).
To scale these results to other weak lensing surveys, equation
(\ref{equaltime}) should be used with a time calibration i.e.
 \begin{equation}
  \label{TT0cal}
  \frac{T}{T_0}= \left(\frac{z_m}{z_{m0}}\right)^{4}
  \left(\frac{A}{A_{0}}\right)
  \left(\frac{D}{D_{0}}\right)^{-2}
  \left(\frac{fov}{fov_{0}}\right)^{-1}.
 \end{equation}
The subscript $0$ refers to the parameters time, median redshift and
area of 
a survey on a telescope with certain diameter and field-of-view.
The scaling applies between surveys with equal number of bands;
for $5$ bands the Canada-France-Hawaii Telescope Legacy Survey
(CFHTLS) can be used, while for $17$ bands COMBO-17 can be used.
For other numbers of bands it can be naively assumed that the time for
a given 
survey scales proportionally with the number of bands so that $T_0
\rightarrow T_0 N_{b0}/ N_{b}$ where $N_{b}$ is the number of
bands in the survey.
\begin{table}
\begin{center}
\begin{tabular}{|l|c|c|}
 \hline
      &   CFHT    &  COMBO-17 \\
 \hline
 D(m) &   3.6      &    2.2 \\
 fov (sq deg.) & 3 &    1    \\
 N (bands) &   5   &    17   \\
 $z_m$  &   1.17   &  0.7    \\
 Area (sq.deg.)& 170 &  1  \\
 T (nights)& 500   &   6   \\
 \hline
\end{tabular}
\caption{Default survey parameters for the 5-band CFHT Legacy
Survey and the 17-band COMBO-17 survey.}
\label{surveytable}
\end{center}
\end{table}
For a flexible survey design the optimal median redshift of
$z_m\approx 1.0$ is approximately insensitive to the 
number of bands, when combined with a Planck
prior (see Figure \ref{zmedian}). If the number of bands is $5$, $9$
or $17$ the appropriate 
line in Figure  \ref{zmedian} then scales proportionally up (and down)
with 
decreased (or increased) areal coverage from $2400$ square
degrees, for a $5$ band survey i.e. $\Delta
w_0(A)=(0.093)(A/2400)^{-1}$. If the number of bands is not shown in
Figure \ref{zmedian} then Figure \ref{optimalsigma} can be used to
find the minimum of the
appropriate $\Delta w_0$ vs. $z_m$ line (at $z_m=0.7$). This can
then be scaled for a differing areal coverage as before.

For a fixed survey of area $A$ and median redshift $z_{m}$
the error in
Figure  \ref{zmedian} for a given median redshift $\Delta w_0(z_{m})$
can be calculated using $\Delta w_0(A)=\Delta
w_0(z_{m})(A/2400)^{-1}$. In scaling between bands a similar  
interpolation between Figure \ref{zmedian} and Figure
\ref{optimalsigma} can be performed.

\subsection{Constraining w(z) at higher redshifts}
\label{Constraining w(z) at higher redshifts}
\subsubsection{Pivot redshifts}
\label{Pivot redshifts}
The parametrization used for the dark energy
equation of state encodes information on both the present day value
of $w$ and its redshift evolution. By placing constraints on both
$w_0$ and $w_a$ a region in the $w(z)$ vs. redshift coordinate system is
constrained. Through the anti-correlation of $w_0$ and $w_a$ this
constraint is minimized at a `pivot redshift', the minimal error at
this redshift being the pivot redshift error. The pivot redshift is
schematically defined by the angle of the ellipse in the ($w_0$,$w_a$)
plane, the error at this redshift being the width of the semi-minor
axis of the ellipse.

\begin{figure}
\psfig{file=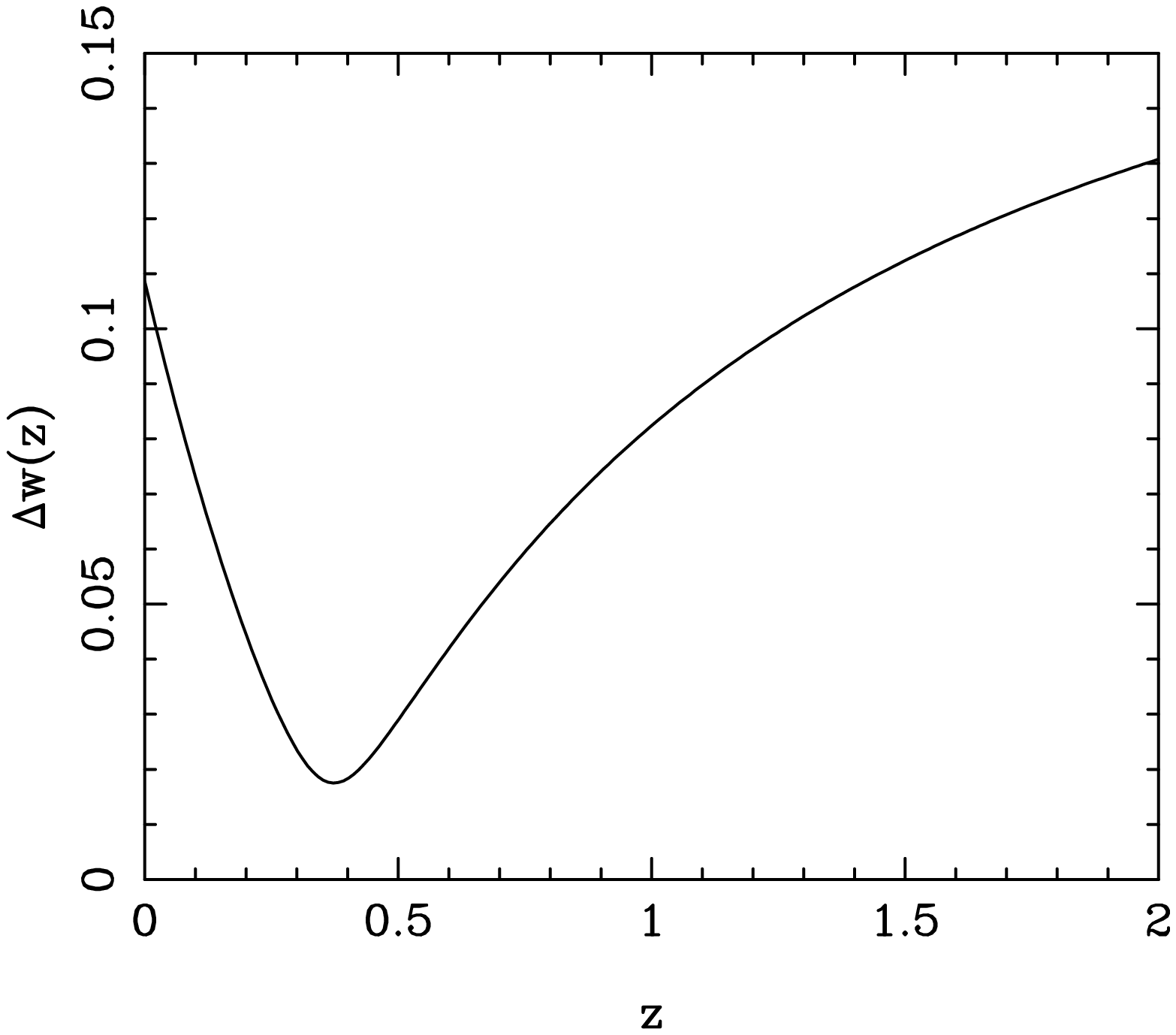,width=\columnwidth,angle=0}
\caption{Marginal error on $w(z)$ combined with a $14$-month Planck
prior. The highest accuracy is achieved at the pivot redshift of
$z=0.373$ with an error of $\Delta w(z=0.373)=0.0175$.}
\label{zerror}
\end{figure}

Figure \ref{zerror} shows the constraint on $w(z)=w_0+w_a(z/(1+z))$
as a function of redshift for our fiducial weak lensing survey. The
highest accuracy on $w(z)$ occurs at the pivot redshift of $z=0.373$
with $\Delta w(z=0.373)=0.0175$.
The pivot redshift of the CMB alone is $z=0.368$
with an error of $\Delta w(z=0.368)=0.0350$, the pivot redshift in
this case is determined by the redshift at which the dark energy
density begins to dominate over the matter density. The pivot
redshift of 3D weak lensing alone is $z=0.208$ with an error of
$\Delta w(z=0.208)=0.2018$ this is determined both by the redshift
at which dark energy becomes dominant and the redshift at which the
lensing signal is maximized.

\begin{figure}
\psfig{file=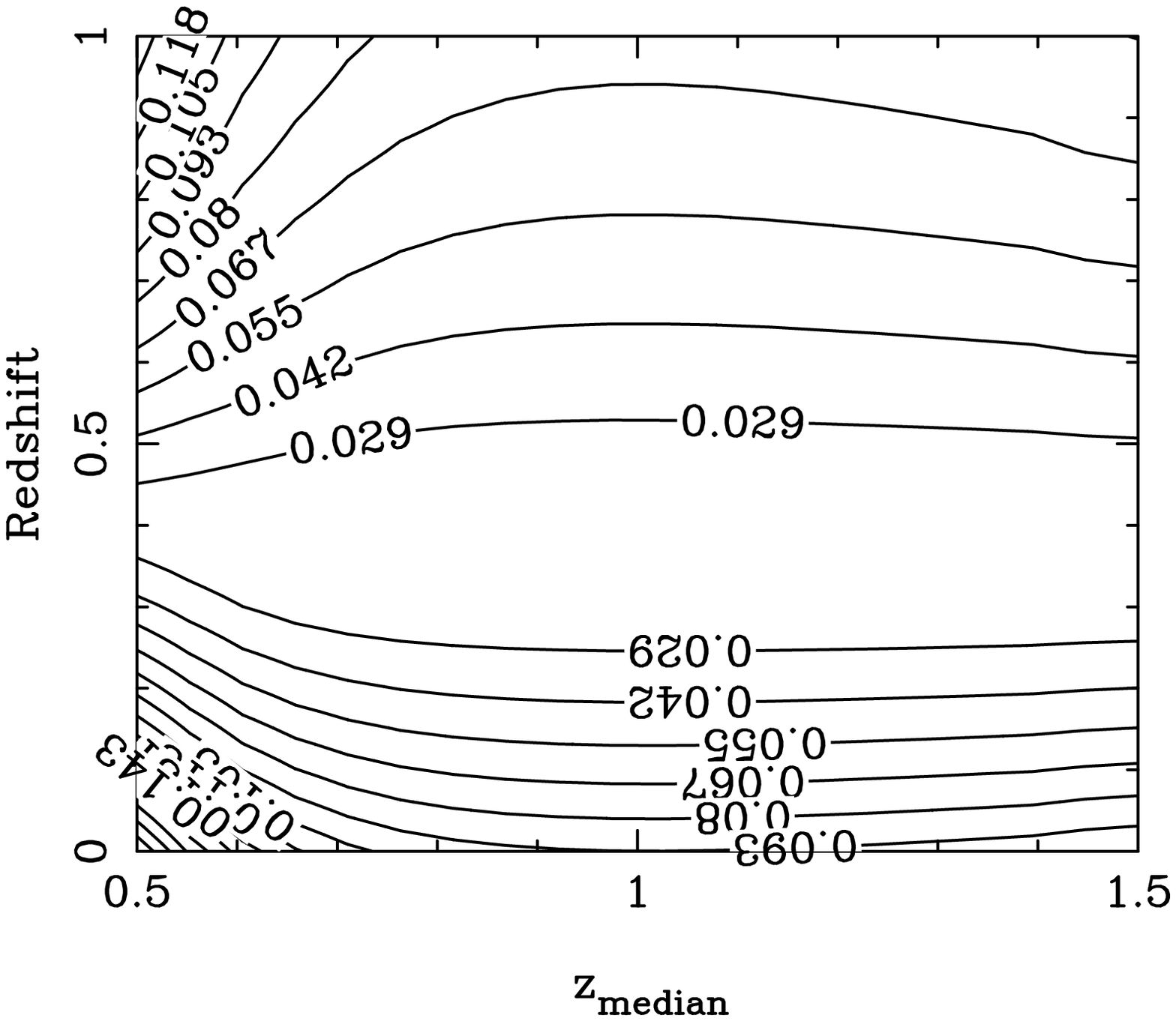,width=\columnwidth,angle=0}
\caption{Marginal error on $w(z)$ combined with a $14$-month Planck
prior as a function of median redshift. The contours are lines of
equal marginal error the values of the contour given on the line.}
\label{zerrorcont}
\end{figure}
Figure \ref{zerrorcont} shows how the error on $w(z)$ varies both with
redshift and the median redshift of the survey. The line in Figure
\ref{zerror} can be found by tracing the $z_m=0.70$ line in Figure
\ref{zerrorcont}, the $5$ band line in Figure \ref{zmedian} can be found by
tracing along the x-axis ($w(z)=w(0)=w_0$) in Figure
\ref{zerrorcont}. There is little sensitivity to the pivot redshift or 
the pivot redshift error on the survey design, this is due to the
intersection of the 3D weak lensing constraint with the $14$-month
Planck constraint remaining the same. This occurs because the pivot
redshift of 3D weak lensing is a property of the cosmological
dependence of the method not the survey design parameters, so that
despite the marginal errors on $w_0$ and $w_a$ varying with the median
redshift the orientation of the lensing ellipse, and hence its
intersection with the $14$-month Planck ellipse, remains the same.

\subsubsection{Figure of Merit}
\label{Figure of Merit}
\begin{figure}
\psfig{file=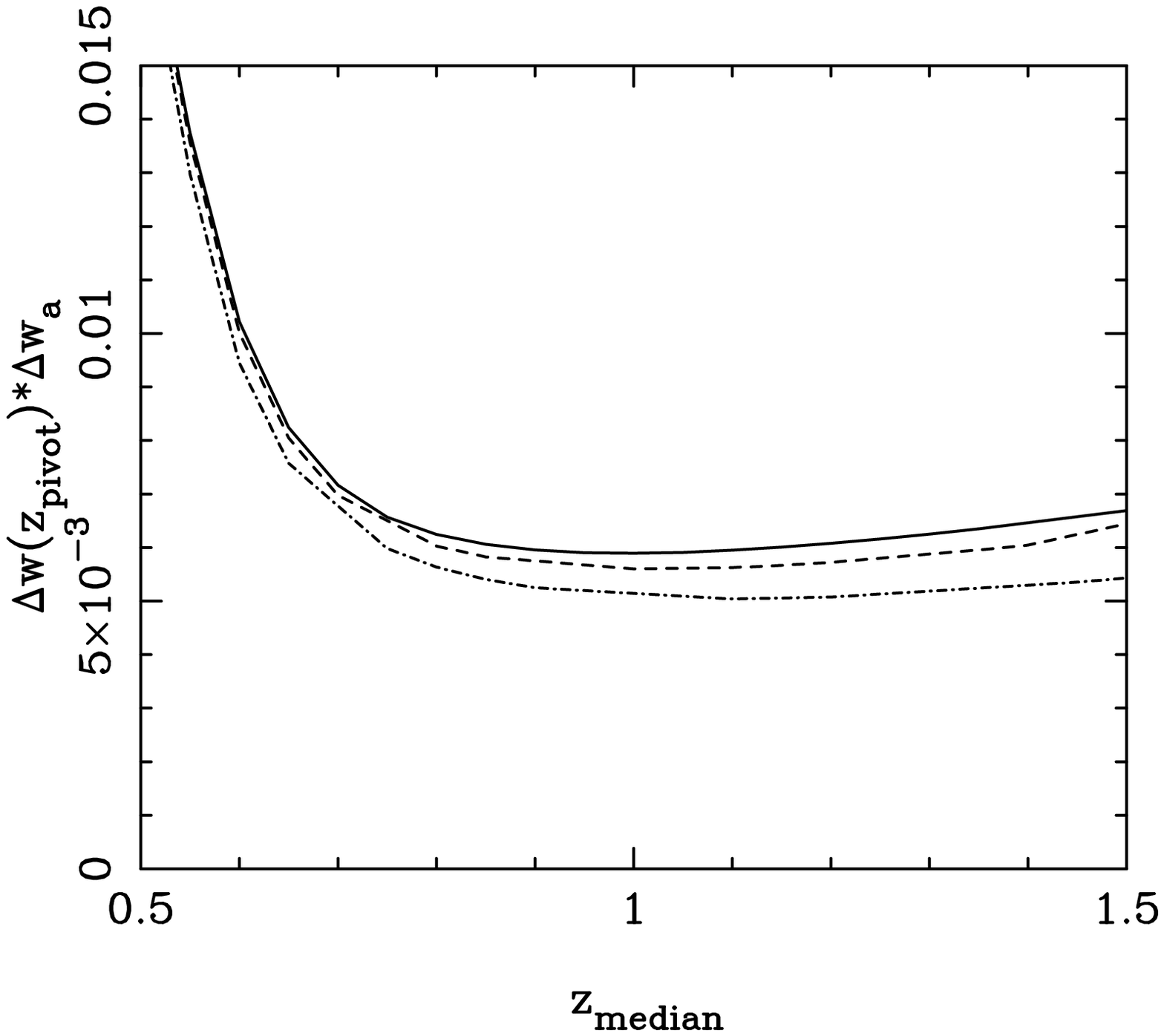,width=\columnwidth,angle=0}
\caption{The figure of merit (product of errors in $w(z_{\rm pivot})$
  and $w_a$) as a function of median redshift for a
  $z_m=0.70$, $A=10,000$ square degree survey including a $14$-month
  Planck prior. Errors in $w_0$ and $w_a$ are
  marginalized over all other parameters. The
  solid line is for $5$-band photometric redshift survey, the dashed
  line for a $9$-band survey and the dot-dashed line for $17$-band
  survey. Note we assume shapes are not measurable beyond $z_{\rm max}=1.5$.}
\label{figozm}
\end{figure}

A `figure of merit' has recently been introduced by the Dark
Energy Task Force (DETF) (2006) (also see Linder, 2006; Taylor et al.,
2006) which 
represents the area of the decorrelated ellipse constrained by a
survey at the pivot redshift and can be written
\begin{equation}
\Delta w(z_{\rm pivot}) *\Delta w_a.
\end{equation}
In reference to Figure \ref{zerror} the figure of merit quantifies
both the minimal pivot redshift error and the redshift range over
which the error in $w(z)$ is small; a wide and deep curve in Figure
\ref{zerror} would have a small figure of merit. Figure \ref{figozm}
shows how the figure of merit for a survey consisting of $600$
nights on a $4$ metre telescope with a $2$ square degree field-of-view
varies with the median redshift of the survey. 
It can be seen that for a $5$, $9$ or $17$ band optical survey the
optimal median redshifts are the same as when optimizing for $w_0$
alone, see Figure \ref{zmedian}. It can be seen that the optimal
median redshift in Figure \ref{figozm} coincides with the widest
point of the inner contour in Figure \ref{zerrorcont}. The Figure of
merit is shown for all considered experiments in Table
\ref{fullresults}.

\section{Synergy of dark energy experiments}
\label{Parameter Forecasts} Here we present the results of comparing
and combining 3D weak lensing with other dark energy probes.
Combining probes, which use different cosmological effects to measure
dark energy either through the growth of structure or geometric
effects, will allow for cross checks to made. These cross checks may
illuminate possible discrepancies between the two effects which
could be important in determining the nature of dark energy. 3D weak
lensing probes both the growth of structure via the matter power
spectrum, and geometry through the lensing effect and the matter
power spectrum.
\begin{figure}
\psfig{file=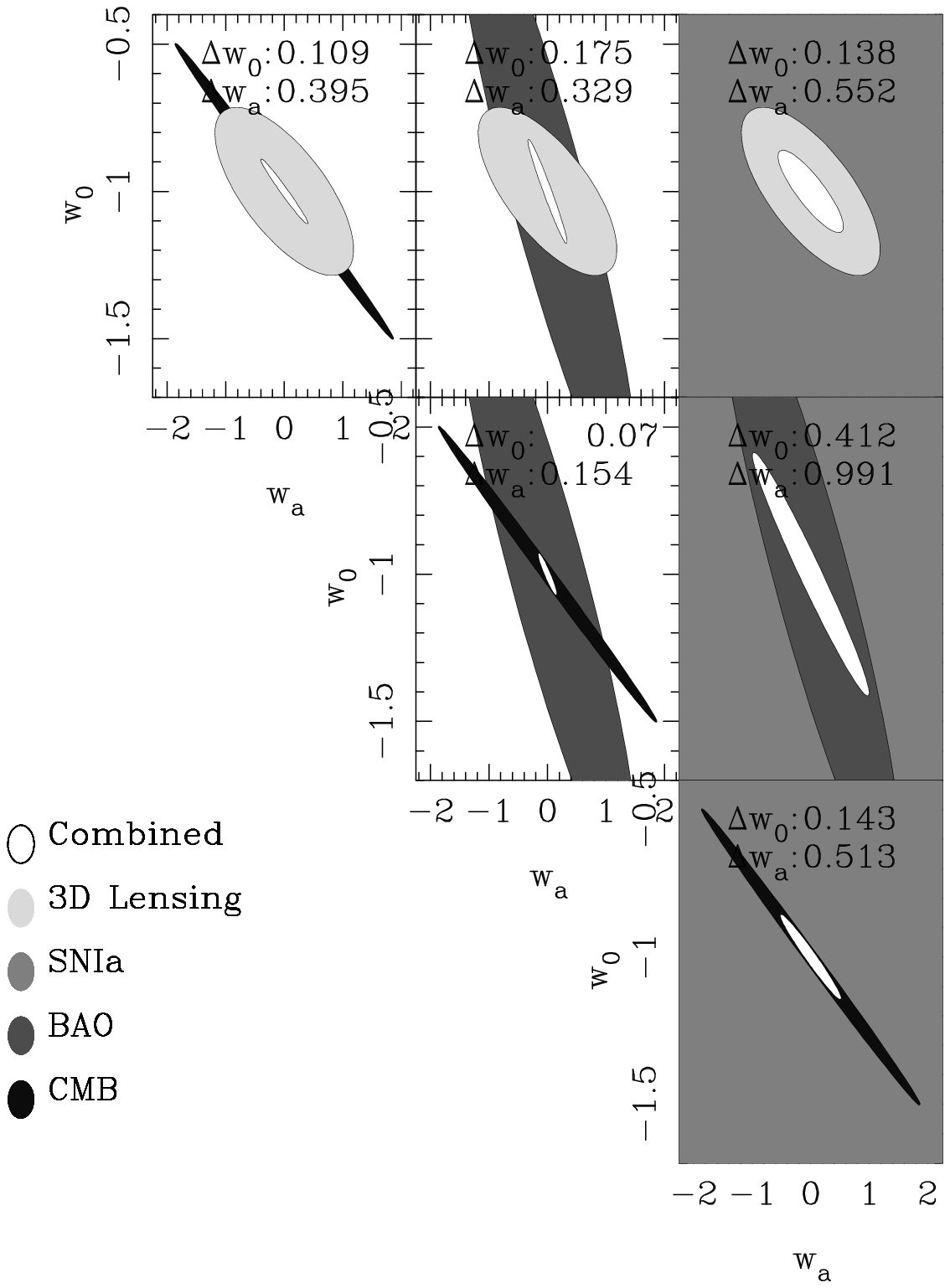,width=\columnwidth,angle=0}
\caption{The combined marginal
  $w_0$, $w_a$ constraints for four individual experiments combined in
  pairs. The experiments 
  are a darkCAM lensing experiment and a CMB 14-month Planck
  experiment, a BAO WFMOS experiment and a SNIa SNAP experiment. The
  dark ellipses in the diagonal panels is the CMB constraint; the
  small, light gray, ellipses along the top row of panels are the
  lensing constraint; the
  second lightest gray ellipses in the right-hand panels are the SNIa
  constraint; the broad darker gray ellipses in the top middle,
  right-hand middle and middle diagonal are the BAO constraint. The
  small white central ellipses are the combined constraints.}  
\label{2Exp}
\end{figure}

\subsection{Comparing and combining with CMB, BAO and SNIA experiments}
\label{Comparing and combining with CMB, BAO and SNIA experiments}
\begin{figure}
\psfig{file=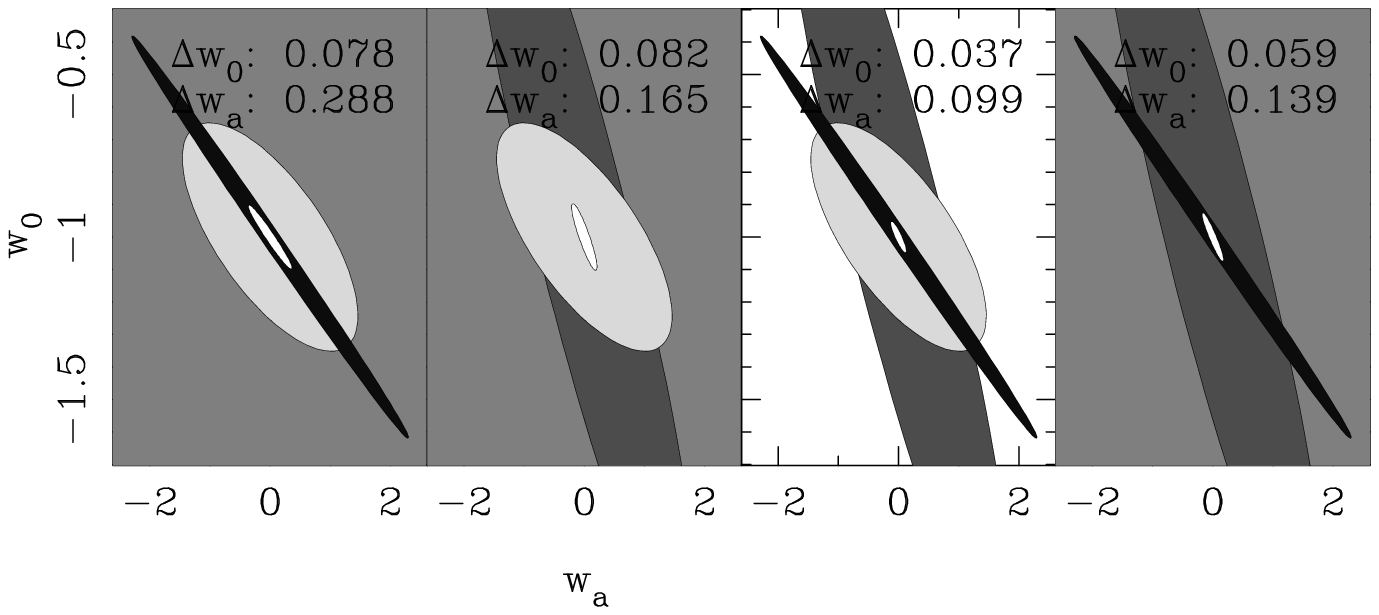,width=\columnwidth,angle=0}
\caption{The combined marginal
$w_0$, $w_a$ constraints for four experiments, combined three at a time. The experiments
are a darkCAM lensing experiment, a CMB 14-month Planck 
experiment, a BAO WFMOS experiment and a SNIa SNAP experiment.}
\label{3Exp}
\end{figure}

\begin{figure}
\begin{multicols}{2}{
\psfig{file=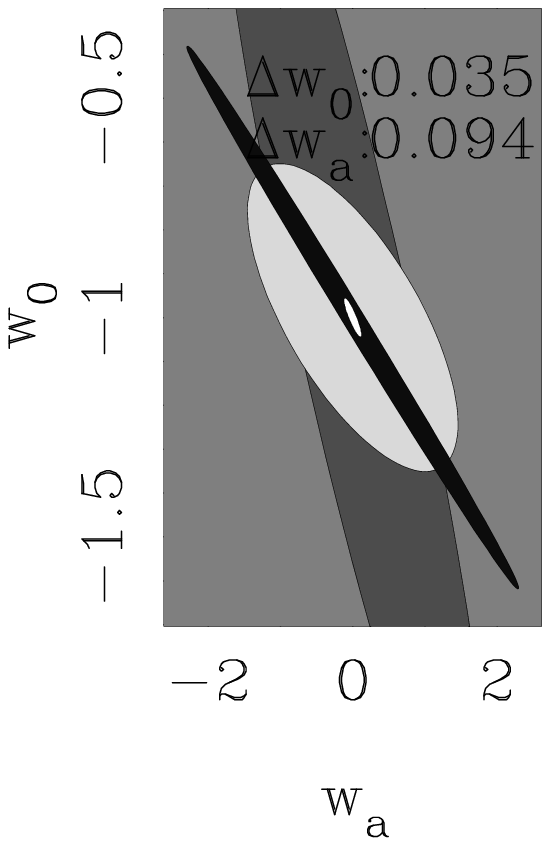,width=0.45\columnwidth,angle=0,clip=}
\psfig{file=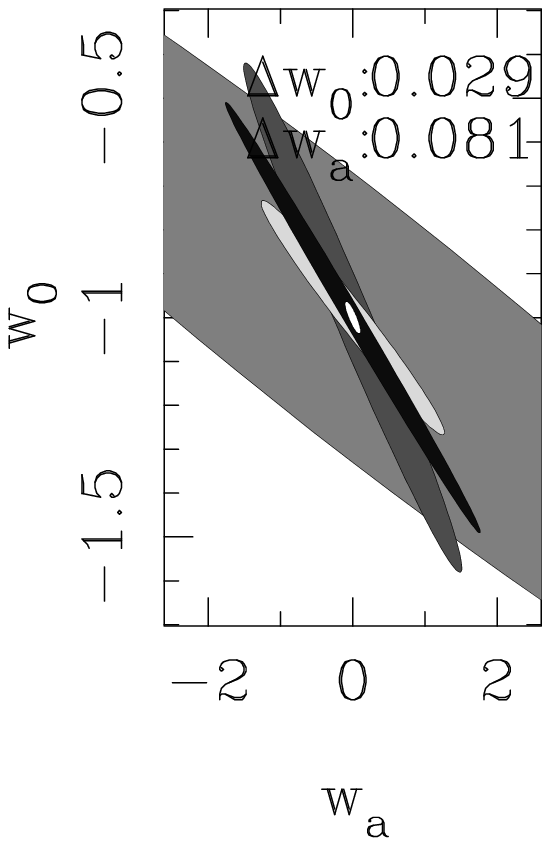,width=0.45\columnwidth,angle=0,clip=}}
\end{multicols}
\caption{The left hand panel shows the combined marginal
  $w_0$, $w_a$ constraints for all four experiments combined allowing
  for fully open models, the right hand panel shows the combined marginal
  $w_0$, $w_a$ constraints for all four experiments combined, with the
  condition $\Omega_m+\Omega_v=1$ enforced. The experiments
  are a darkCAM lensing experiment and a CMB 14-month Planck
  experiment, a BAO WFMOS experiment and a SNIa SNAP experiment.}
\label{4Exp}
\end{figure}

We consider the CMB, SNIa and BAO experiments as described in Section
\ref{Combining with other dark energy experiments}. In Figures
\ref{2Exp}, \ref{3Exp} and \ref{4Exp} the dark
thin ellipse is the CMB constraint; the small lightest gray
ellipse is the lensing constraint; the darker gray, almost
vertical, broad ellipse is the BAO constraint; the very broad
lighter gray ellipse is the SNIa constraint. 
 
Figure \ref{2Exp} shows the combined two parameter $1$-$\sigma$
(68.3\%) contours for all the possible pairs of experiments considered.
In comparison with the other methods the 3D weak lensing constraint
provides the smallest independent constraint. In combination the
marginal errors so not vary largely between the different pairs. The
BAO and CMB pair combination provides the smallest marginal errors
through the unique degeneracy of the BAO ellipse providing a small
intersection with the CMB. The SNIa constraint alone is poor
although in combination with the other dark energy probes does
provide an improvement on the marginal errors through the
intersection of the constraints. The SNIa is also a purely geometric
test, so that the combination with 3D weak lensing would provide an
important cross check.

We combine combinations of three experiments in Figure \ref{3Exp}.
It can be seen, by comparing with Figure \ref{2Exp}, that in adding
further information from another experiment improves all the
combined marginal errors. The largest improvements are gained by
adding 3D weak lensing, or the CMB, to the pair combinations that do
not include these probes. The smallest marginal errors are achieved
by combining 3D weak lensing with the CMB and BAO. In combining 3D
weak lensing with SNIa and BAO the dark energy constraints are
comparable, or better than, each pair combination that includes the
CMB constraint.

Finally the combination of all four of the dark energy probes is
shown in Figure \ref{4Exp}. By adding 3D weak lensing to the three
experiment combination of CMB, BAO and SNIa the marginalized
constraints are improved by a factor of $2$. The marginalized
constraints in the combination of all four of the probes considered
are 
\begin{eqnarray}
\Delta w_0&=&0.035\nonumber\\
\Delta w_a&=&0.094,
\end{eqnarray}
with a pivot redshift
error of 
\begin{equation}
\Delta w(z=0.43)=0.0147.
\end{equation}
Reducing the maximum $\ell$ to $2000$ increases these errors by
approximately $0.01$ to
$\Delta w_0=0.045$ and $\Delta w_a=0.105$.

\subsection{Complementary figures of merit and pivot redshifts}
\label{Complementary figures of merit and pivot redshifts}
\begin{figure}
\psfig{file=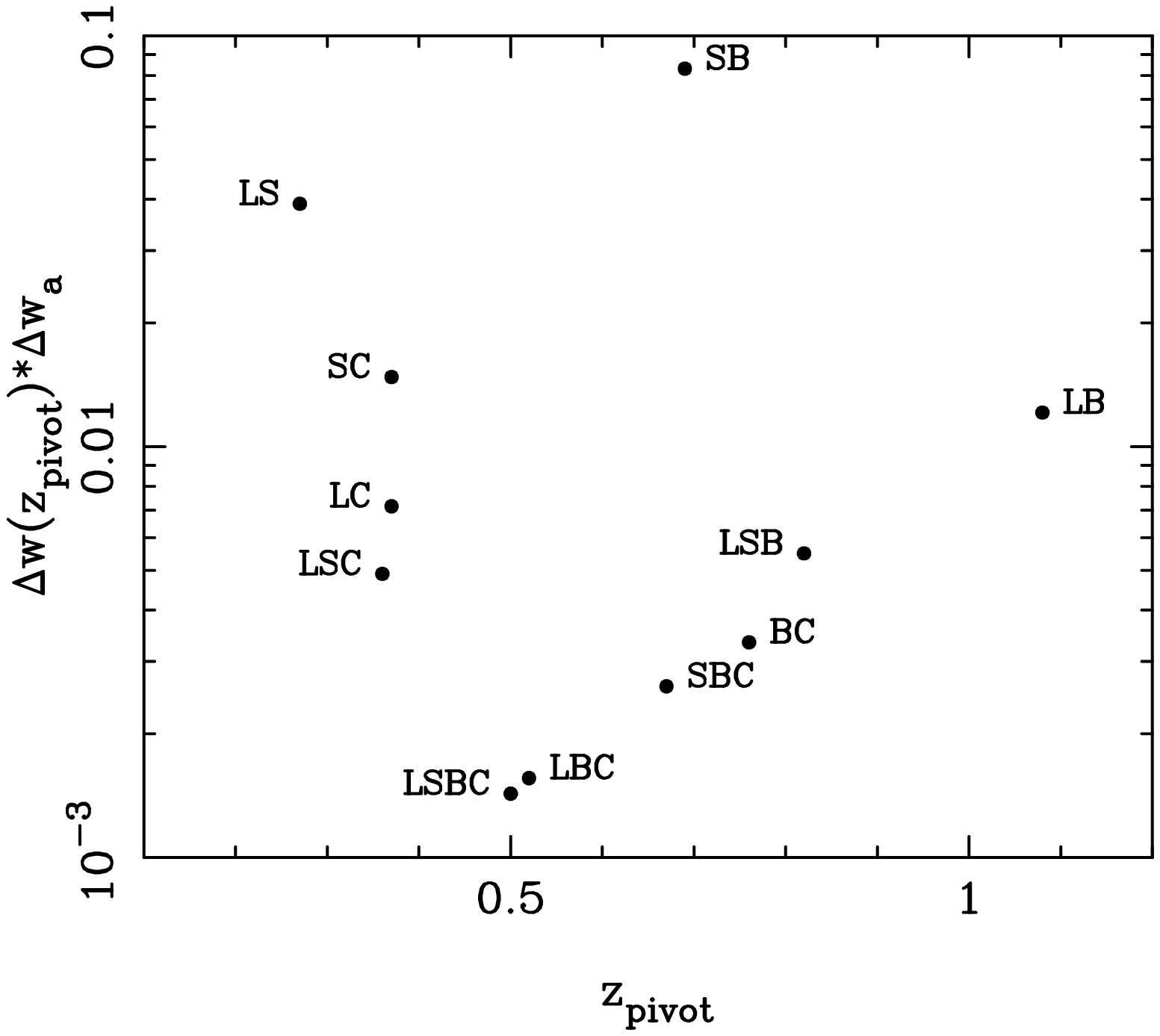,width=\columnwidth,angle=0}
\caption{The figure of merit and pivot redshift for various
  experimental combinations. The combinations are labeled as
  L=3D weak Lensing, B=BAO, S=SNIa, C=CMB.
  Combinations of letters represent combinations of experiments.}
\label{FOMPIV}
\end{figure}

An illustrative way to present the information of pivot redshifts and
the figure of merit of a dark energy probe, or combination of
different probes, is to show how the figure of merit and the pivot
redshift compare. We show this in Figure \ref{FOMPIV} by plotting the
figure of merit for all the possible combinations of experiments
against the pivot redshift of the combined constraint.
In general the more experiments that are added in combination the
smaller the figure of merit becomes. As more experiments are added
in combination the pivot redshift converges to a single value. In
combination with other experiments the BAO constraint creates a high
pivot redshift, this is due to the redshift of the nearest bin at
$z=1$. The 3D 
weak lensing constraint in combination creates a low $z\approx 0.4$ pivot
redshift; this is due to the lensing pivot redshift dominating which
is a symptom of the lensing signal maximizing at around that
redshift. It can be seen from Figure \ref{FOMPIV} that there exists
combinations, 
for example 3D weak lensing with the CMB (LC) and 3D weak lensing with
BAO and SNIa (LSB), that have similar figures of merit but very
different pivot redshifts. In 
using combinations such as these the $w(z)$ evolution could be
constrained to a high degree over a large redshift range.
\begin{table*}
\begin{tabular}{|l|c|c|c|c|c|c|c|c|}
\hline
Survey& Area sqdeg &$z_{median}$&$N_{bands}$ & $\Delta w_0$&$\Delta
w_a$&$z_{\rm pivot}$&$\Delta w(z_{\rm pivot})$&$\Delta w(z_{\rm pivot})$$\Delta
w_a$\\
\hline
\hline
{\bf Lensing}&$$&$$&$$&$$&$$&$$&\\
\hline
darkCAM + Planck&$10000$&$0.7$&$5$&$0.1082$&$0.3966$&$0.3681$&$0.0175$&$0.0069$\\
\hline
darkCAM + BAO darkCAM&$10000$&$0.7$&$5$&$0.2764$&$1.1207$&$0.2086$&$0.2004$&$0.0418$\\
\hline
darkCAM, 9 bands +
Planck&$10000$&$0.7$&$9$&$0.1072$&$0.3895$&$0.3733$&$0.0173$&$0.0067$\\
\hline
SNAP Lensing + SNIa +
Planck&$1000$&$1.38$&$5$&$0.0579$&$0.2322$&$0.3247$&$0.0112$&$0.0026$\\
\hline
All-Sky Space +
Planck&$40000$&$1.0$&$9$&$0.0101$&$0.0406$&$0.3047$&$0.0342$&$0.0014$\\
\hline
darkCAM+Planck+BAO+SNIa&$10000$&$0.7$&$5$&$0.0350$&$0.0944$&$0.5011$&$0.0151$&$0.0014$\\
\hline
VST-KIDS+WMAP4&$1400$&$0.6$&$5$&$0.3405$&$1.0818$&$0.4378$&$0.0862$&$0.0933$\\
\hline
CFHTLS(Wide)+WMAP4&$170$&$1.17$&$5$&$0.2541$&$0.8145$&$0.4275$&$0.0711$&$0.0579$\\
\hline \hline
{\bf CMB}&$$&$$&$$&$$&$$&$$&\\
\hline
4-year WMAP&$$&$$&$$&$2.060$&$3.612$&$1.18$&$0.758$&$2.7379$\\
\hline
14-Month Planck&$$&$$&$$&$0.501$&$1.873$&$0.367$&$0.035$&$0.0655$\\
\hline \hline
{\bf BAO}&$$&$$&$$&$$&$$&\\
\hline
BAO WFMOS+Planck&$2000$&$1.0$&$$&$0.070$&$0.154$&$0.78$&$0.019$&$0.0029$\\
\hline
\hline
{\bf SNIa}&$$&$$&$$&$$&$$&$$&$$\\
\hline
  SNIa SNAP+Planck&$$&$$& $$ & $0.142$ & $0.513$&$0.37$&$0.028$&$0.0144$\\
\hline
\end{tabular}
\caption{Expected marginal errors on cosmological parameters from 3D
analysis of proposed weak lensing surveys. Note here $9$ bands refers
to $5$ optical bands plus $4$ infrared.}
\label{fullresults}
\end{table*}

\subsection{Future lensing surveys}
\label{ss:Future} There are a number of current and planned imaging
surveys for weak lensing which could be analyzed in 3D.  The
surveys vary in depth, areal coverage and number of bands, and
illustrative errors are shown in the Table \ref{fullresults}. Errors
are marginalized. The surveys considered are: the Canada France
Hawaii Legacy Survey (CFHTLS; Semboloni et al., 2006) which is ongoing;
the VST (VLT Survey Telescope) public survey KIDS; SNAP
(Supernova/Acceleration Probe; Aldering, 2005), and darkCAM on a $4$ metre
telescope.  
We show the errors achievable with darkCAM combined with various
different experiments. BAO darkCAM refers to using the photometric
redshifts from darkCAM to measure BAO. VST-KIDS and CFHTLS have been
combined with a $4$ year WMAP prior as Planck will not be
contemporary with these surveys. Here 9 bands refers to a $5$ band
optical survey 
with $4$ infrared bands as discussed in Section \ref{Optical and
  Infrared surveys}. 

\subsection{The effect of changing the fiducial dark energy model}
\begin{figure}
\psfig{figure=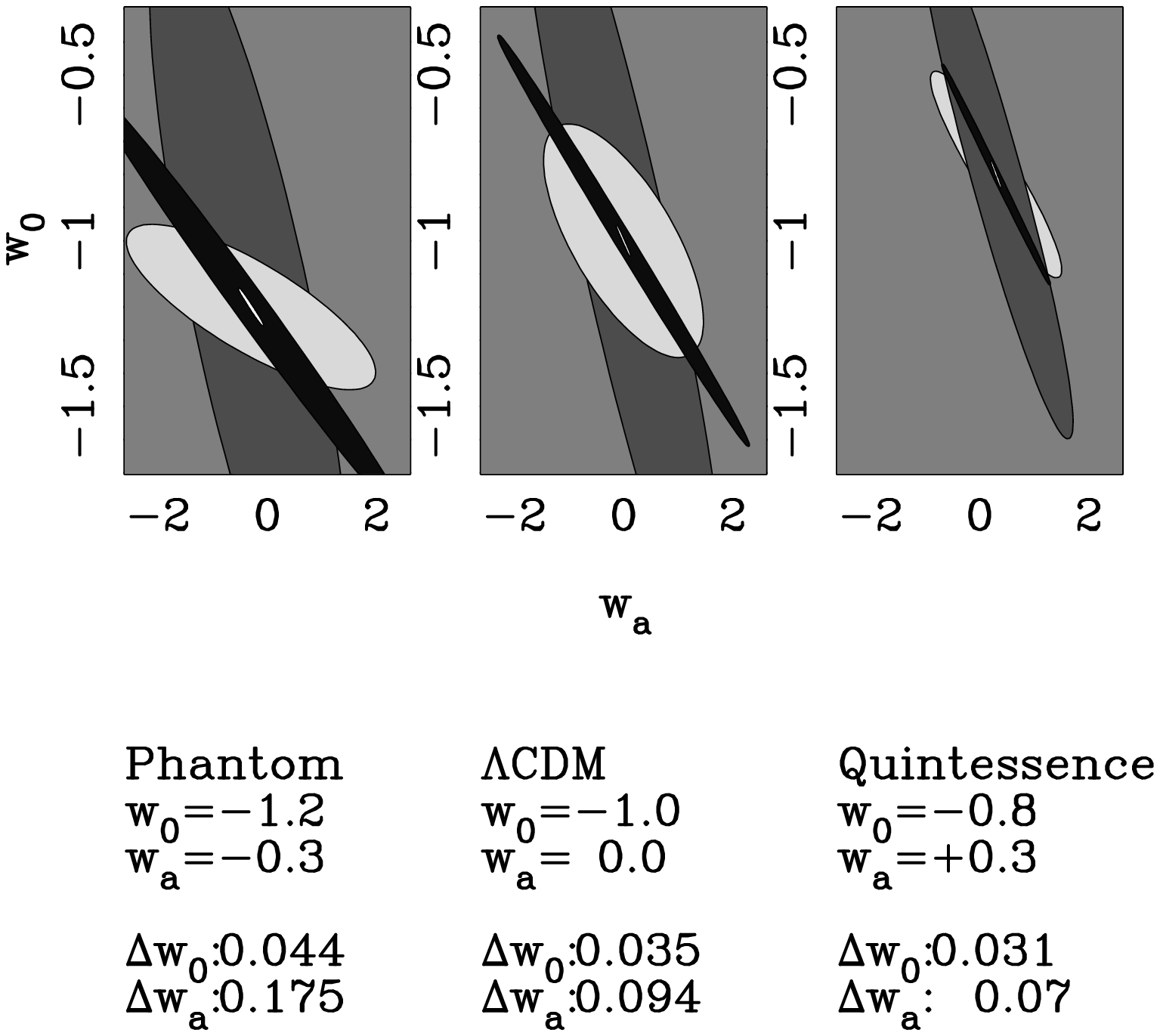,width=\columnwidth,angle=0,clip=}
\caption{The two parameter $1$-$\sigma$ (68.3\%) contours for various
  assumed fiducial dark energy models, for a  10,000
  square degree
  survey to a median depth $z_m=0.7$, with a 14-year Planck prior, a BAO
  WFMOS prior and a SNIa SNAP prior. The errors quoted are the one
  parameter $1$-$\sigma$ marginal errors on $w_0$ and $w_a$. The dark
  thin ellipse is the CMB constraint; the small lightest gray
  ellipse is the lensing constraint; the darker gray, almost
  vertical, broad ellipse is the BAO constraint; the very broad
  light gray ellipse is the SNIa constraint. The small white central
  ellipses are the combined constraints.} 
  \label{DEScen}
\end{figure}

Here we investigate the effect of the assumed fiducial dark energy 
cosmology on the marginal errors from our Fisher matrix
calculations. The assumed cosmology has been a cosmological constant
model with $w_0=-1$ and $w_a=0$. Figure \ref{DEScen} shows the two
parameter $1$-$\sigma$ (68.3\%) contours for various dark energy
models in the ($w_0$, $w_a$) plane fully marginalized over other
parameters. We consider two extreme examples, just allowable from
current constraints:  a SUGRA (Super Gravity) model proposed by
Weller \& Albrecht (2002) represented by $w_0=-0.8$ and $w_a=+0.3$;
and a phantom model proposed by Caldwell et al. (2003) with
$w_0=-1.2$ and $w_a=-0.3$. Despite the marginal errors from the dark energy experiments alone
changing, the combined marginal error on $w_0$ is largely unaffected
by the assumed dark energy model. The main difference is occurs on
the error on $w_a$ which increases for all methods as its value
becomes more negative. This is due to the fact that a negative $w_a$
represents a dark energy scenario in which the dark energy density
was less in the past (increasing into the future); so that the
effect of dark energy on the expansion rate on observed galaxies (in
the past) is less in these scenarios (and similarly the opposite
effect for a positive $w_a$). For the range of $w_0$ and $w_a$
allowed by current constraints the marginalized errors presented
here should be robust to the actual nature of dark energy.

\subsection{The effect of assuming flatness}
Here we present the effect of assuming flatness in the parameter error 
estimation i.e. $\Omega_v=1-\Omega_m$. Figure \ref{4Exp} shows 
the two parameter $1$-$\sigma$ (68.3\% contour) constraints for all four 
dark energy probes considered, see Section \ref{Combining with other
  dark energy experiments},  
in the ($w_0$, $w_a$) plane 
with the assumption that the Universe is flat. The CMB $14$-month Planck 
constraint does not considerably improve because the CMB puts a strong
constraint  
on the overall geometry of the Universe, through the position of the
first acoustic peak. The improvement in the SNIa
constraint is  
most evident, since this dark energy probe is very sensitive to the
overall geometry  
through the Hubble parameter (this is in agreement with Linder,
2005). The 3D weak lensing and BAO constraints also considerably 
improve, with the overall combined errors on the dark energy equation
of state 
parameters being $\Delta w_0=0.029$ and $\Delta w_a=0.089$, a factor
of $1.2$ less than 
the constraints considering fully open models. In assuming flatness
the different dark energy probes still have unique and complimentary
degeneracies in the ($w_0$, $w_a$) plane. The reduction in predicted
errors, especially in the SNIa, 3D weak lensing and BAO experiments,
show that  
the assumption of flatness can have an affect on parameter error 
estimation (this is in agreement with the affect of this assumption
on weak lensing tomographic methods, see for example Knox, Song \&
Zhan, 2006). Given that  
some proposed dark energy models rely on modifications to the
Friedmann equation 
in non-flat geometries it is prudent to calculate predicted parameter
errors using fully  
open geometries.

\section{Conclusions}
\label{Conclusions} In this paper we have presented a 3D weak lensing
spectral method suitable for high-$\ell$ studies, and investigated how
well 3D weak 
lensing surveys could determine the equation of state of dark
energy.  The accuracy which could be achieved if systematic errors
can be controlled is impressively high, provided the surveys are
analyzed in 3D: marginal statistical errors of $\Delta w_0=0.108$, on
the current value of $w\equiv p/(\rho c^2)$, and its evolution $w_a$
constrained to $\Delta w_a=0.397$ are possible with a 10,000 square
degree survey in 5 bands to a median source depth of $z_m=0.7$. At a
pivot redshift of $z=0.37$ such an experiment could constrain $w(z)$
to  $\Delta w(z=0.37)=0.0175$. Such a survey is possible with darkCAM,
in conjunction with data from the 
Planck satellite. Even without Planck, the accuracy from 3D weak
lensing alone is still impressively high, and better than any other
dark energy probe considered on its own. The fact that the physics
of 3D weak lensing is well-understood, combined with the small
statistical error forecasts, makes 3D weak lensing a formidable
prospect for advancing cosmology in the next decade. The errors on
$w$ are comparable to, but a little better than, predictions from
tomography (Hu and Jain, 2004; Ishak, 2005). The constraints on $w(z)$ at the
pivot redshift and the figure of merit, $\Delta w(z_{\rm pivot})*\Delta w_a$,
of the experiments considered were also discussed.

We have investigated optimizing a wide field survey to measure the
equation of state parameters $w_0$ and $w_a$ and found an optimal
survey strategy of $z_m=1.0$ covering $2400$ square degrees for a
$5$ optical band survey. We found that increasing the number of
optical bands to $9$ or $17$ makes little difference to the marginal
errors when the 3D weak lensing result is combined with a Planck
prior. The effect of including infrared bands in a wide field survey
was investigated by varying the photometric redshift error, it was
found that adding $4$ infrared bands to a $5$ band optical survey
improves the marginal constraints on $w_0$ slightly from $\Delta w_0=0.108$
to $\Delta w_0= 0.097$.

Three alternative dark energy probes were considered: a Planck CMB
experiment; a WFMOS BAO experiment and a SNAP SNIa experiment. All
possible combinations of experiments were considered and the figure
of merit and pivot redshifts of the combinations shown. In such a
competitive environment 3D weak lensing places strong constraints on
the dark energy parameters and in combination with other experiments
provides a unique degeneracy in the ($w_0$,$w_a$) plane which is
manifest as a strong constraint at a particular pivot redshift.

We have addressed the issues of biased photometric redshift estimates (e.g.
Ma, Hu and Huterer, 2005) and show that the method is relatively insensitive to
this. We also investigated the effect that a sample of outliers, with
poor photometric redshift estimates, would have on the predicted
marginal errors. The effect of outliers on the marginal error of $w_0$
is small although the way in which such a sample is treated is important. 

We have not considered errors due to the intrinsic alignment of
galaxies (Heavens, Refregier and Heymans, 2000; Croft and Metzler,
2000; Catelan, Kamionkowski and Blandford, 2001; Crittenden et al.,
2001; Jing, 2002), as
these may be reduced to a negligible level by removing pairs which
are close in photometric redshifts (Heymans and Heavens, 2003; King
and Schneider, 2002). This
procedure has already been demonstrated in the analysis of the
COMBO-17 data (Heymans et al., 2004). We have also not addressed other issues
of systematics, such as optical distortions, or possible alignment of
foreground galaxies with shear (Hirata and Seljak, 2004), which may be reduced
using techniques such as template fitting (King, 2005). Nevertheless, the
fact that the statistical errors are very small is very encouraging.
Clearly to achieve the accuracies quoted here is going to be a
formidable challenge for control of systematics, but at least the
statistical error forecasts are small enough that the promise of
accurate measurement of the equation of state of dark energy may be
realized.

\section{Acknowledgments}
We would like to thank David Bacon and Graca Rocha for
helpful discussions, Masahiro Takada for discussions concerning the
CMB and BAO predictions, Chris Wolf for discussions concerning the
photometric redshift error formalism. TDK acknowledges a PPARC studentship.


\begin{thebibliography}{}

\bibitem{} Abbott L., Schaefer R., 1986, ApJ, 308, 546

\bibitem{} Aldering, G., 2005, NewAR, 49, 346

\bibitem{} Bacon D., Refregier A., \& Ellis R., 2000, MNRAS, 318, 625

\bibitem{} Ballinger W., {Heavens} {A.}, {Taylor} {A.}, 1995, MNRAS,
  276, 59

\bibitem{} {Bartelmann} {M.} and {Schneider} P., 2001, Phys. Rep., 340, 291

\bibitem{} Bassett B., et al., 2005, A\&G, 46e, 26

\bibitem{} {Baugh} {C.} and {Efstathiou} G., 1993, MNRAS, 265, 145

\bibitem{} Bennett C. et al., 2003, ApJS, 148, 1

\bibitem{} Bernstein G., 2005, astro-ph/0503276

\bibitem{} Bernstein G., Jain B., 2004, ApJ, 600, 17

\bibitem{} Blake C., Bridle S., 2005, MNRAS, 363, 1329 

\bibitem{} Brown M. et al., MNRAS, 314, 100

\bibitem{} {Castro} {P}, {Heavens} {A.}, {Kitching} {T.}, 2005, Phys Rev
D72, 023516 (astroph/0503479)

\bibitem{} Catelan P., Kamionkowski M., Blandford R., 2001, MNRAS, 320, 7

\bibitem{} Chevallier M., Polarski D., 2001, Int. J. Mod. Phy. D., 10, 213

\bibitem{} {Colless} {M.}, et al, 2001, MNRAS, 328, 1039

\bibitem{} Crittenden R. et al., 2001, ApJ, 559, 552

\bibitem{} Croft R. and Metzler C., 2002, ApJ, 545, 561

\bibitem{} Croft R. et al. 2002, ApJ, 581, 20

\bibitem{} DETF, 2006, http://www.nsf.gov/mps/ast/detf.jsp

\bibitem{} {Fisher} {K.}, {Scharf} {C.}, {Lahav} {O.}, 1994, MNRAS, 266, 219

\bibitem{} {Gnedin} {N.}, {Hamilton} {A.}, 2002, MNRAS, 334, 107

\bibitem{} {Heavens} {A.F.}, 2003, MNRAS, 343, 1327

\bibitem{} {Heavens} {A.}, Refregier A., {Heymans} {C.}, 2000, MNRAS,
319,649

\bibitem{} {Heavens} {A.}, {Taylor} {A.}, 1995, MNRAS, 275, 483

\bibitem{} Heymans C., Brown M., Heavens A., Meisenheimer K., Taylor A. Wolf C.
2004, MNRAS, 347, 895

\bibitem{} Heymans C., Heavens A., 2003, MNRAS, 339, 711

\bibitem{} Hirata C. \& Seljak U. 2004, Phys. Rev. D204, 3056

\bibitem{} Hoekstra H., Yee H., Gladders M., 2002, ApJ, 577, 595

\bibitem{} Hoekstra H. et al., 2006, astroph, 0511089

\bibitem{} {Hu} {W.}, 1999, ApJ, 522, 21

\bibitem{} {Hu} {W.}, 2002, Phys. Rev. D66, 3515

\bibitem{} Hu W. \& Jain, B., 2004, Phys. Rev. D, 70, 043009

\bibitem{} Hu W. \& Tegmark M., 1999, ApJ, 514, L65

\bibitem{} Huterer D., 2002, Phys. Rev. D, 65, 063001

\bibitem{} Huterer D. et al.,  2005, astroph/0506030

\bibitem{} Ishak M., Hirata C., McDonald P. \& Seljak U., 2004, Phys. Rev.
D, 69, 083514

\bibitem{} Ishak M., 2005, astro-ph/0501594

\bibitem{} {Jain} {B.}, Taylor A., 2003, Phys. Rev. Lett., 9, 1302

\bibitem{} Jarvis M. et al., 2003, AJ, 125, 1014

\bibitem{} Jarvis M., Jain B., Bernstein G. \& Dolney D., 2005, astro-ph/0502243

\bibitem{} Jing Y.P., 2002, MNRAS, 335, 89

\bibitem{} Jungman G., Kamionkowski M., Kosowsky A., Spergel D., 1996,
  PhRvD, 54, 1332

\bibitem{} Kaiser N., Wilson G., \& Luppino G., 2000, astro-ph/0003338

\bibitem{} King L., 2005, A\&A, 441, 47

\bibitem{} King L., Schneider P., 2002, A\& A, 396, 411

\bibitem{} Kim A., et al., 2004, MNRAS, 347, 909

\bibitem{} Knox L., Scoccimarro R., Dodelson S., 1998, PhRvL, 81, 2004

\bibitem{} Knox L., Song Y., Zhan H., 2006, astro-ph/0605536

\bibitem{} Kuo C.L. et al., 2004, ApJ, 600, 32

\bibitem{} Lamarre J. et al., 2003, NewAR, 47, 1017

\bibitem{} Linder E.,  2005, APh, 24, 391

\bibitem{} Linder E., 2003, Phys. Rev. Lett., 90, 091301

\bibitem{} Linder E., Jenkins A., 2003, MNRAS, 346, 573

\bibitem{} Ma Z., Hu W., Huterer D., 2005, astroph/0506614

\bibitem{} {Peebles} {P.}, 1980, The Large-Structure of the Universe,
Princeton University Press, Princeton

\bibitem{} {Pearson} {T.}, {et al.}, 2003, ApJ, 591, 556

\bibitem{} {Percival} {W.}, {et al.}, 2001, MNRAS, 327, 1297

\bibitem{} {Percival} {W.}, {et al.}, 2004, MNRAS, 353, 1201

\bibitem{} Ratra B., Peebles P., 1988, Phys. Rev. D37, 3406

\bibitem{} Refregier, A., 2003\ ARAA, 41, 645

\bibitem{} Rhodes J. et al., 2004, ApJ, 605, 29

\bibitem{} Riess A.G. et al.\ 2001, ApJ, 560, 49

\bibitem{} {Santos} {M.}, {et al.}, 2003, MNRAS, 341, 623

\bibitem{} {Seljak} {U.}, Zaldarriaga M., 1996, ApJ, 469, 437

\bibitem{} Semboloni E. et al., 2006, A\&A, 452, 51S

\bibitem{} Simon P., King L., Schneider P., 2004, A\& A, 417, 873

\bibitem{} Smith R. et al.\ 2003, MNRAS, 341, 1311

\bibitem{} Song Y. \& Knox L. 2004, Phys. Rev. D, 70, 063510

\bibitem{} {Spergel} D. et al, 2003, ApJS, 148, 175

\bibitem{} {Spergel} D. et al, 2006, astroph, 0603449

\bibitem{} {Tadros} H. et al, 1995, MNRAS, 305, 527

\bibitem{} Takada, M. \& Jain, B. 2004, MNRAS, 348, 897

\bibitem{} {Takada} {M.}, {White} {M.}, 2004, ApJL, 601, 1

\bibitem{} Taylor A., 2005, ADS, pdus, confE, 28, {http://www.noao.edu/meetings/subaru}

\bibitem{} Taylor A., Kitching T., Bacon D., Heavens A., 2006, astro-ph/0606416 

\bibitem{} Tegmark M., Taylor A., {Heavens} {A.}, 1997, ApJ, 480, 22

\bibitem{} Van Waerbeke, L. et al. 2000, A\&A 358, 30

\bibitem{} Verde L. et al., 2002, MNRAS, 335, 432

\bibitem{} Wester W., 2005, ASPC, 339, 152

\bibitem{} White, M. 2004, Astropart. Phys., 22, 211

\bibitem{} Wittman D., Tyson J., Kirkman, D., Dell'Antonio I., \&
    Bernstein, G. 2000, Nature, 405, 143

\bibitem{} Wolf C., et al., 2001, A\&A, 377, 442

\bibitem{} Wolf C., et al., 2003, A\&A, 401, 73

\bibitem{} Wolf C., et al., 2004, A\&A, 421, 913

\bibitem{} {Y\`eche} C., et al., 2006, A\&A, 448, 831

\bibitem{} Zaldarriaga L.,  Seljak U., 2000, ApJS, 129, 431

\bibitem{} Zhan H. \& Knox L. 2004, Astrophys. J., 616, L75

\end{thebibliography}
\end{document}